\documentclass[manuscript]{aastex}

\usepackage{emulateapj5}

\slugcomment{Accepted to ApJ}

\shorttitle{T Dwarf Binaries}
\shortauthors{Burgasser et al.}

\begin{document}

\title{Binarity in Brown Dwarfs:
T Dwarf Binaries Discovered with the Hubble Space Telescope WPFC2}

\author{
Adam J.\ Burgasser\altaffilmark{1,2}, J.\ Davy Kirkpatrick\altaffilmark{3},
I.\ Neill Reid\altaffilmark{4}, Michael E.\ Brown\altaffilmark{5}, Cherie L.\
Miskey\altaffilmark{6}, and John E.\ Gizis\altaffilmark{7}}

\altaffiltext{1}{Division of Astronomy and Astrophysics,
University of California at Los Angeles, Los Angeles, CA,
90095-1562; adam@astro.ucla.edu} \altaffiltext{2}{Hubble Fellow}
\altaffiltext{3}{Infrared Processing and Analysis Center, M/S
100-22, California Institute of Technology, Pasadena, CA 91125;
davy@ipac.caltech.edu}
\altaffiltext{4}{Space Telescope Science Institute, 3700 San
Martin Drive, Baltimore, MD 21218; inr@stsci.edu}
\altaffiltext{5}{Division of Geological and
Planetary Sciences, M/S 105-21, California Institute of
Technology, Pasadena, California 91125; mbrown@gps.caltech.edu}
\altaffiltext{6}{Institute for Astrophysics and Computational
Sciences, Department of Physics, Catholic University of America,
Washington, DC 20064; and NASA Goddard Space Flight Center,
Greenbelt, MD 20771; miskey@iacs.gsfc.nasa.gov}
\altaffiltext{7}{Department of Physics and Astronomy, University
of Delaware, Newark, DE 19716; gizis@udel.edu}

\begin{abstract}
We present the discovery of two T dwarf binaries, 2MASS
1225$-$2739AB and 2MASS 1534$-$2952AB, identified in a sample of
ten T dwarfs imaged with the Hubble Space Telescope Wide Field
Planetary Camera 2. Companionship is established by the uniquely
red F814W$-$F1042M colors of the binary components, caused by
heavily pressure-broadened K I absorption centered at 7665 \& 7699
{\AA}.  The separations of the two binary systems are
0{\farcs}282$\pm$0{\farcs}005 and 0{\farcs}065$\pm$0{\farcs}007,
implying projected separations of 3.17$\pm$0.14 and 1.0$\pm$0.3
AU, respectively. These close separations are similar to those
found in previous brown dwarf binary searches, and permit
orbital mapping over the coming decade.  2MASS 1225$-$2739AB
has a substantially fainter secondary, with $\Delta$M$_{F814W}$ =
1.59$\pm$0.04 and $\Delta$M$_{F1042M}$ = 1.05$\pm$0.03; this
system is likely composed of a T6 primary and T8 secondary with
mass ratio 0.7--0.8.
The observed binary fraction of our HST sample,
20$^{+17}_{-7}$\%, is consistent with results obtained for late-M
and L field dwarfs, and implies a bias-corrected binary fraction
of 9$^{+15}_{-4}$\% for $a \gtrsim 1$ AU and $q \gtrsim 0.4$,
significantly lower than the binary fractions of F--G and early-type
M dwarf stars.
Neither of the
T binaries have separations $a \gtrsim 10$ AU, consistent with
results from other brown dwarf binary searches.  Using the
statistical models of Weinberg, Shapiro, \& Wasserman, we conclude
that tidal disruption by passing stars or Giant Molecular Clouds,
which limits the extent of wide stellar binaries, plays no role in
eliminating wide brown dwarf binaries, implying either disruption
very early in the formation process (ages $\lesssim 1-10$ Myr) or
a formation mechanism which precludes such systems.  We find that
the maximum binary separation in the brown dwarf regime appears to
scale as M$_{total}^2$, a possible clue to the physical mechanism
which restricts wide substellar systems.

\end{abstract}

\keywords{stars: binaries: visual ---
stars: formation ---
stars: fundamental parameters ---
stars: individual (2MASS J12171110$-$0311131,
2MASS J12255432$-$2739466, 2MASS J15344984$-$2952274) ---
stars: low mass, brown dwarfs
}

\section{Introduction}

T dwarfs are low-temperature (T$_{eff}$ $\lesssim$ 1300--1500 K)
brown dwarfs exhibiting distinct absorption bands of CH$_4$ in the
near-infrared H- and K-bands \citep{kir99,me02,geb02}. They are
distinguished from warmer L-type \citep{kir99,mrt99} and M-type
brown dwarfs by the presence of the CH$_4$ bands, in addition to
significant H$_2$O and collision-induced (CIA) H$_2$ absorption in
the near-infrared.  These molecular features, combined with
pressure-broadened K I and Na I absorption at red-optical
wavelengths \citep{tsu99,bur00,lie00a}, force the emergent
spectral energy distributions of T dwarfs to peak around 1
$\micron$. Since the discovery of the prototype of this class,
Gliese 229B \citep{nak95,opp95}, over 30 T dwarfs have been
identified in wide-field surveys such as the Two Micron All Sky
Survey \citep[hereafter 2MASS; Skrutskie et al.\
1997]{me99,me00c,me02d,me02} and the Sloan Digital Sky Survey
\citep[hereafter SDSS; York et al.\
2000]{str99,tsv00,leg00,geb02}, deep narrow-field surveys
\citep{cub99,liu02,zap02}, and as companions to nearby stars
\citep{nak95,me00a,els01}.

For both stars and brown dwarfs, multiplicity is one of the
fundamental properties that probes formation processes. Multiple
systems are common among main sequence stars, with roughly 60\% of
solar-type stellar systems found to be comprised of two or more
components \citep{abt76,abt87,duq91}. In contrast, only 32--42\%
of field M dwarf systems are multiple \citep{hen90,fis92,rei97}.
Recent investigations of late-M and L field dwarf samples yield
even smaller multiplicity fractions, only 20-30\%
\citep{koe99,rei01a,clo02}. Furthermore, while the separation
distribution of F-M stellar pairs appears to be broad (ranging
from approximately 0.1 AU to 0.1 pc) and unimodal (peaking around
3-30 AU), all late-M and L dwarf binaries identified to date have
apparent separations $a < 0{\farcs}6$ and projected separations
$a_{proj} < 15$ AU \citep{mrt99c,koe99,rei01a,leg01,clo02}.
Ejection models for brown dwarf formation \citep{rpt01,bat02} have
been proposed to explain this dearth of widely-separated, low-mass
dwarf pairs, which may be the result of nature (i.e., inherent in
the formation process itself) or nurture (i.e., due to dynamical
scattering).

In this article, we present imaging results for the first T dwarf
binary search sample, undertaken using the Hubble Space Telescope
(HST) Wide Field Planetary Camera 2 \citep[hereafter
WPFC2]{bir01}. In $\S$2 we describe the observations and image
reduction techniques, and identify closely-separated sources near
two of our T dwarf targets.  In $\S$3 we describe our photometric
analysis, presenting aperture photometry and colors for our
complete sample.  The colors of the individual components of the
2MASS 1225$-$2739\footnote{Throughout the main text, we abbreviate
object names to 2MASS hhmm$\pm$ddmm; full designations are given
in Table 1.} and 2MASS 1534$-$2952 pairs confirm their
companionship. We discuss our PSF fitting in $\S$4, by which we
derive rigorous flux ratios and separations for the two T
binaries, and quantify our search limits. In $\S$5 we discuss
individual targets in detail, including a possible faint companion
to 2MASS 1217$-$0311.  In $\S$6 we analyze binary statistics for
field L and T dwarfs, and compare to stellar samples.  Finally, we
discuss our results in light of brown dwarf binary formation and
destruction processes in $\S$7, and conclude that the small
separations of these systems are not due to disruptive encounters
with stars and Giant Molecular Clouds (GMCs) in the Galaxy, but
are more likely imposed early on in the formation process.

\section{Observations\label{sec:observations}}

We imaged a sample of ten T dwarfs identified in a
magnitude-limited search of the 2MASS database
\citep{me99,me00a,me00c,me02} in the WPFC2 F814W (${\lambda}_c =
7940${\AA}) and F1042M (${\lambda}_c = 10201${\AA}) filters during
HST Cycle 9.  A log of observations is given in Table 1. The F814W
and F1042M filters were chosen to sample the red wing of the
pressure-broadened K I doublet, as the strength of this feature
allows late-type L and T dwarfs to stand out from nearly all
background sources in red optical colors (e.g., I--z; Leggett et
al. 2000). Each object was centered on the PC chip and observed
twice (to allow for cosmic-ray subtraction) in both filters for
total exposure times ranging from 2000 to 2600 sec.

Images were reduced by standard pipeline processing, which
includes analog-to-digital correction, removal of the bias
pedestal, subtraction of bias and dark frames, and division by an
appropriate flat field image. No correction to shutter shading was
required due to the long exposure times. The images were then
combined using a cosmic ray rejection routine developed at NASA
Goddard, based on an improved version of the CR{\_}REJECT routine
written by R.\ S.\ Hill. Bad pixels identified both in the data
quality arrays and the cosmic ray rejection routine were replaced
by the mean of neighboring pixels to produce the final, cleaned
image.  We note that a reflection arc from an offset bright star
appears on the F814W
PC chip images of 2MASS 2356$-$1553, caused by non-optimal
baffling of this detector \citep{bir01}.  This reflection leads to
a slightly higher background in the vicinity of the source but
does not affect our background-subtracted photometry ($\S$3.1). No other
residual or reflection artifacts are seen in the data.

Sections of the reduced PC
images 2{\farcs}3 on a side around each of the primary
targets are shown in Figure 1.
North/east orientations are indicated by arrows.  We
immediately identify two closely-separated objects near the position
of 2MASS 1225$-$2739
in both the F814W and F1042M images, with the western component appearing to
be fainter at both bands.  2MASS 1534$-$2952 is slightly elongated
along a north/south axis as compared to both the other T dwarf targets and
other sources in the 2MASS 1534$-$2952 PC field, as is shown in more
detail in Figure 2.  We attribute
this elongation to a marginally resolved pair of point sources
($\S$4.2).  All of the
other T dwarf targets appear to be single point sources at the
spatial resolution of the PC chip (0$\farcs$046).

\section{Photometry}

\subsection{Aperture Photometry}

Sources on all four WPFC2 chips were initially identified with the
IRAF\footnote{IRAF is distributed by the National Optical
Astronomy Observatories, which are operated by the Association of
Universities for Research in Astronomy, Inc., under cooperative
agreement with the National Science Foundation.} DAOFIND routine,
and then confirmed by visual inspection. We extracted aperture
photometry for each source using the IRAF PHOT routine, using 2,
3, and 5 pixel apertures, corresponding to 0{\farcs}09,
0{\farcs}14, and 0{\farcs}23 on the PC chip and 0{\farcs}20,
0{\farcs}30, and 0{\farcs}50 on the WF chips. Background values
were determined using a centroid algorithm in a 15 pixel annulus
10 pixels from each source. Individual aperture corrections were
then measured for each single star (neglecting galaxies and
residual cosmic rays) by integrating their radial profiles to 20
pixels.  Because our fields were generally at high galactic
latitudes, source densities were low, and many fields
(particularly those in the F1042M filter) had few or no stellar
sources on a particular chip.  Hence, a mean set of aperture
corrections were derived from all point sources in each chip and
filter; these values are given in Table 2. After applying the
aperture corrections, flux values were corrected for geometric
distortion and charge-transfer efficiency (CTE), the latter by the
prescription of \citet{whi99}. Because of the long exposure times
and high backgrounds, typical CTE corrections generally did not
exceed 20\%.  No corrections for instrument contamination were
made, as they are exceedingly small in these red filters
\citep{bir01}. Synthetic flux zeropoints from \citet{bag97} were
used to convert the corrected magnitudes to the WPFC2 photometric
system.  A final source list was compiled by requiring detections
in both filters and positional coincidence within 1 pixel; this
constraint eliminated most residual cosmic rays (however, see
$\S$5.2).

Because aperture corrections were made for point-source radial
profiles, extended sources such as galaxies or close doubles were
readily identified by significant differences in derived
photometry depending on aperture size. This effect verified the
slight elongation of 2MASS 1534$-$2952. All other T dwarf targets
(including the two close sources in the 2MASS 1225$-$2739 field)
have photometry in each aperture consistent with the formal
uncertainties.  We adopt the 3-pixel aperture magnitudes for point
sources (optimizing the signal-to-noise ratio), except for the
second source in the 2MASS 1225$-$2739 field, where we select a
2-pixel aperture to minimize the contribution of the brighter
nearby source; and extended sources, including 2MASS 1534$-$2952,
where we select 5-pixel aperture magnitudes to minimize aperture
corrections. HST WPFC2 and 2MASS J-band magnitudes and colors for
our T dwarf targets and Gliese 229B\footnote{J-band photometry for
Gliese 229B is on the UKIRT system \citep{cas92}, which is similar
to the 2MASS photometric system \citep{car01,dah02}.}
\citep{gol98,leg99} are listed in Table 3.

\subsection{T Dwarf Colors}

Figure 3 plots F814W magnitude versus F814W$-$F1042M color for all
sources identified in the ten WPFC2 datasets, along with data for
Gliese 229B. Single point sources and target objects are plotted
as solid circles, while extended sources (i.e., galaxies) are
plotted as open circles. Primary T dwarf targets are individually
labelled, all of which are 2--3 magnitudes redder than the
background stellar and galactic sources, again due to the red wing
of the pressure-broadened K I doublet. Both sources at the
position of 2MASS 1225$-$2739 lie at red colors, implying that
both are T dwarfs.  Based on the estimated surface density of T
dwarfs detectable by 2MASS, 8.4$\times$10$^{-4}$ deg$^{-2}$
\citep{me02}, which we extrapolate to a limiting J magnitude of
17 (the apparent J magnitude of 2MASS 1225$-$2739B; see below),
the probability of two relatively bright T dwarfs randomly
lying within $\sim$18$\arcsec$ of each other (the approximate
search radius on the PC chip) is 3$\times$10$^{-7}$. We therefore
confidently claim companionship for these two objects based on
their proximity and unique colors. By the same argument, the two
sources at the position of 2MASS 1534$-$2352 are also companion T
dwarfs, based on the red color of their combined light. Hereafter,
we refer to these two systems as 2MASS 1225$-$2739AB and 2MASS
1534$-$2352AB.

Figure 4 plots the F814W$-$F1042M versus F814W$-$J color-color
diagram for the observed T dwarfs and Gliese 229B.  Note that
the colors of single targets follow a fairly linear trend:
\begin{equation}
[F814W-J] = (3.1{\pm}0.3) + (0.46{\pm}0.07){\times}[F814W-F1042M].
\end{equation}
Because 2MASS 1225$-$2739 is unresolved by 2MASS, we determined J-band
component magnitudes from the combined light magnitude, J = 15.22$\pm$0.05,
and the J-band flux ratio,
\begin{equation}
{\Delta}J = {\Delta}F814W - {\Delta}[F814W-J] = 1.35{\pm}0.08,
\end{equation}
using Eqn.\ 1 and the photometry listed in Table 3. The F814W$-$J
colors for these two objects both lie $\sim$0.15 mag below the
linear fit traced by the single stars, but are consistent within
the photometric uncertainties.
The combined light F814W$-$J color of 2MASS 1534$-$2952AB is also
below the single star locus, but in this case it is probably
because the object is marginally resolved in the WPFC2 images. On
the other hand, 2MASS 1217$-$0311 is slightly redder in F814W$-$J
color than expected, although by no more than 2$\sigma$. Both
F814W$-$F1042M and F814W$-$J colors are generally redder for the
later-type T dwarfs, with the former being particularly sensitive
to spectral type. One notable exception is the T6.5 emission-line
dwarf 2MASS 1237+6526 \citep{me00b}, which has the reddest
F814W$-$F1042M color in the sample (see $\S$5.3). On the other
hand, F1042M$-$J colors generally decrease for later spectral
types, likely due to increased H$_2$O and CH$_4$ absorption around
1.25 $\micron$ \citep{me02}.

\section{PSF Fitting}

\subsection{Technique}

In order to derive separations and flux ratios for our two T dwarf
binaries, and search for faint companions around the other target
sources, we performed point spread function (PSF) subtraction on
all of our primary targets. Our technique was as follows: first,
we extracted subimages of all apparently single point sources from
the PC chip images of all ten datasets, a total of 22 sources in
F814W and 11 in F1042M. These included some of the target objects,
although care was taken to exclude any point sources with bad
pixels near the source peak. We then subtracted two-dimensional
Gaussian fits to the PSFs from the images; typical residuals were
consistently $\lesssim 10$\% of the original source peak. Finally,
we averaged these Gaussian-subtracted images, scaled by the fit
maximum, to produce a single PSF residual image for each filter.

For each of our target sources,
we searched for faint companions using an iterative ${\chi}^2$ reduction
routine.  Model images were constructed by combining two PSF residual
images with two Gaussian
surfaces having the same FWHM as the PSF fits described above,
each scaled to separate component fluxes.
For 2MASS 1225$-$2739AB and 2MASS 1534$-$2952AB, initial guesses
were based on the approximate
positions and flux ratios from the aperture photometry (we assumed
2MASS 1534$-$2952AB to be separated by 1 pixel along each axis as an
initial guess).  Our
routine then iteratively searched for the optimal solution to the
primary position, secondary position, primary flux, and secondary flux,
in that order, by shifting the component positions in steps of 0.1 pixels
and scaling the
fluxes in steps of 1\% (0.01 mag).  If the secondary flux was scaled
below 1 count or separations below 0.5 pixels were reached, then the
object was considered a single point source.  Otherwise, the routine
derived separations, position angles, and flux ratios for the
optimal binary solution.

For all of the apparently single targets, we followed up this process by
fitting a single PSF residual plus Gaussian to the image and
then searching by eye for any
obvious counterparts.  We then used the same binary search routine
for each image
with 20 random companion initial positions.  If no
companion brighter than the S/N = 7 detection limits
(approximately 25.5 mag at F814W and 19.9 mag at F1042M; Table 5)
was found, the primary target
was assumed to be single.

\subsection{Results}

Convergent binary solutions were obtained for both 2MASS
1225$-$2739AB and 2MASS 1534$-$2952AB in both filters. Figure 5
shows the original and PSF-subtracted images for these pairs.
Residuals from the subtraction were less than 4--5\% of the peak
flux, at the level of 8--10 (F814W) and 2--3 (F1042M) times the
background noise.
In both cases, smaller residuals were obtained for fits to two
sources rather than a single source.  This validates the
duplicity of 2MASS 1534$-$2952AB, whose separation
(0$\farcs$065$\pm$0$\farcs$007) is smaller
than both the diffraction limit (0$\farcs$08 at F814W, 0$\farcs$11
at F1042M) and Nyquist sampling limit (2$\times$0$\farcs$046 = 0$\farcs$09)
of the instrument.  We are able to overcome the former because our
technique resolves even significant overlap of two PSFs,
particularly when they have nearly equal brightness.  The latter
constraint is
overcome by using a PSF generated from multiple measurements,
allowing us to subsample below the Nyquist limit.
Table 4 lists the derived binary parameters for these
systems; no other sources can be seen in the subtracted
images. PSF fitting of the other primary T dwarf targets revealed
only one potential faint companion to 2MASS 1217$-$0311, detected
at F1042M only.  This possible detection, which may simply be a residual
cosmic ray, is discussed in detail in $\S$5.2.  No other
companions were identified around any of the other target objects
within 1$\farcs$1, and no other faint objects with T dwarf-like
colors were identified in any of the WPFC2 images.

To obtain proper calibration and determine
the uncertainties of our results,
we ran the algorithm
described above on 20,000 simulated binaries constructed from
the F814W and F1042M images of 2MASS 0559$-$1404.  These test images sampled
a range of separations 1--15 pixels
(0$\farcs$05--0$\farcs$69), all orientations, and
flux ratios $\Delta$M = 0--7 mag.
Once processed through the PSF fitting routine,
those test cases having output separations within 0.5 pixels
and corrected flux ratios within 0.2 mag of
the input values were considered recovered binaries.  Corrections
and uncertainties to both positions and flux ratios were then
determined separately for 2MASS 1225$-$2739AB and 2MASS 1534$-$2952AB
in each filter, using only those recovered test cases having similar
input separations ($a > 0\farcs23$ and $a < 0\farcs14$, respectively)
and flux ratios ($1 < {\Delta}M < 2.5$ and $0 < {\Delta}M < 1$, respectively).
Typical flux ratio corrections
were approximately $-$0.10 mag (i.e., shifting the secondary
to brighter magnitudes) with 1$\sigma$ uncertainties of 0.04 and 0.3
mag for 2MASS 1225$-$2739AB and 2MASS 1534$-$2952AB, respectively;
separation 1$\sigma$ uncertainties were 0.12 and 0.15 pixels
(0$\farcs$005 and 0$\farcs$007), respectively, translating
into position angle uncertainties of 7$\degr$ and 9$\degr$.
The values listed in Table 4 reflect these corrections
and uncertainties.

\subsection{Search Limits}

Our calibration simulations allowed us to derive limiting detection magnitudes
as a function of separation, as shown in Figure 6.  Around 2MASS
0559$-$1404, faint secondaries ($\Delta$M $\gtrsim$ 3) were
generally missed at separations closer than 0$\farcs$15, while
{$\Delta$}M $\approx$ 5.5 (6) and 4.5 (5) could be obtained for
wide separations at the 95\% (50\%) confidence level at F814W and
F1042M, respectively.  In general, only near-equal-magnitude
companions with $a \lesssim
0{\farcs}09$ could be recovered better than 50\% of the time,
as is the case for 2MASS 1534$-$2952AB.

For $a \gtrsim 0{\farcs}4$, S/N = 7 limits (Table 5) yield the
maximum sensitivity for faint companions, ranging from $\Delta$M =
4.3--6.9 at F814W to $\Delta$M = 2.9--4.9 at F1042M.  We can
convert these values to mass ratio ($q \equiv$ M$_2$/M$_1$) limits
using a mass-luminosity power law from \citet{bur01}, L $\propto$
M$^{2.64}$, and assuming for simplicity coevality and negligible
variation in bolometric corrections over the sample\footnote{The
fact that F814W$-$F1042M varies by 1 mag over this sample implies
that this is in general not true, particularly for very cool
companions; however, this assumption is suitable for a rough
limit.}, such that
\begin{equation}
q_{lim} = 10^{\frac{-{\Delta}M_{lim}}{6.6}}.
\end{equation}
These values are listed in Table 5, and range from 1.0--0.4.
Overall, our sample is complete to $q \gtrsim 0.4$ for $a \gtrsim
4$ AU (assuming a mean distance of 10 pc), with less sensitivity
for small mass ratios to separations approaching $a \approx 1$ AU.

\section{Individual Targets}

\subsection{Binaries}

2MASS 1225$-$2739AB is clearly resolved into two unequal-magnitude
components in our HST images.  The colors of these objects are
significantly different, with the fainter companion having an
F814W$-$F1042M color similar to the T8 Gliese 570D \citep{me00a},
while the color of the primary is consistent with the spectral
type of the combined system, T6. The magnitude ratios of this pair,
$\Delta$M$_{F814W}$ = 1.59$\pm$0.04 and $\Delta$M$_{F1042M}$ =
1.05$\pm$0.03, are greater than the absolute magnitude ratios of
Gliese 229B (T6.5) and Gliese 570D, $\Delta$M$_{F814W}$ =
1.01$\pm$0.10 and $\Delta$M$_{F1042M}$ = 0.87$\pm$0.12, consistent
with 2MASS 1225$-$2739A being earlier than type T6.5. Based on
these colors, we speculate that this system is comprised of a T6
and T8 pair, which should be confirmed with spatially-resolved
spectroscopy. No parallax has been measured for this system yet,
but the spectrophotometric distance of the secondary, compared to
Gliese 570D, is $d^B_{F814W}$ = $d^B_{F1042M}$ = 11.1 pc.  At
J-band, $d^A_{J}$ = 10.8 and $d^B_{J}$ = 11.9, based on the
absolute J-band magnitudes of the T6 SDSS 1624+0029 \citep[M$_J$ =
15.33$\pm$0.07]{dah02} and Gliese 570D \citep[M$_J$ =
16.47$\pm$0.05]{me00a}. All of these distance estimates
combined yield a mean $d^{AB}$ =
11.2$\pm$0.5 pc and projected physical separation $a$ =
3.17$\pm$0.14 AU; note that the uncertainties do not represent
probable scatter in the absolute magnitude/spectral type relation,
not yet adequately measured for the T dwarfs.
Adopting T$_{eff}$ $\approx$ 1000 and 800 K for
the two components \citep{me02}, assuming coevality, and using the
models of \citet{bur97}, we can derive component masses for ages
of 0.5, 1.0, and 5.0 Gyr, as listed in Table 6.  The derived mass
ratio of this system is $q$ = 0.7--0.8, depending on its age.
Assuming that, on average, the semimajor axis of a binary system
$<a_{sm}> = 1.26<a>$ \citep{fis92}, we estimate orbital periods of
24--40 yr; hence, significant orbital motion (${\Delta}{\phi}$
$\sim$ 10$\degr$) could be detectable in this system on a yearly
basis.

2MASS 1534$-$2952AB is a more marginally resolved system,
suggesting that it is much more closely-separated than 2MASS
1225$-$2739AB. The flux ratios for this system,
$\Delta$M$_{F814W}$ = 0.5$\pm$0.3 and $\Delta$M$_{F1042M}$ =
0.2$\pm$0.3, are much smaller than the spread in absolute
magnitudes for mid-type T dwarfs; e.g., $\Delta$M$_{F814W}$ =
2.06$\pm$0.06 and $\Delta$M$_{F1042M}$ = 1.36$\pm$0.08 between the
T5 2MASS 0559$-$1404 and the T6.5 Gliese 229B.
Hence, we assume that this system is comprised of two nearly
equal-mass components with spectral types T5.5 and T$_{eff}$
$\approx$ 1100 K.  Again, no parallax measurement has been made for
this system.  Based on the absolute J-band magnitudes of 2MASS
0559$-$1404 and SDSS 1624+0029, we estimate a spectrophotometric
distance of $d_{J}$ = 16$\pm$5 pc, the uncertainty dominated by
the poor absolute magnitude constraints for mid-type T dwarfs.
Combined with the measured separation, this implies $a$ =
1.0$\pm$0.3 A.U. Based on mass estimates as derived above (Table
6), we estimate orbital periods of 4--6.5 yr, making this system
an excellent target for rapid orbital mapping; however, its very
close separation may hamper these measurements, and only
significant orbital motion (${\Delta}{\phi} \gtrsim 30\degr$) may
be detectable.

\subsection{A Potential Binary}

PSF subtraction of the F1042M image of 2MASS 1217$-$0311 reveal
a faint companion 0$\farcs$209$\pm$0$\farcs$006 from the target
source at position angle 74$\pm$7$\degr$.  The corrected flux
ratio of the secondary, $\Delta$M$_{F1042M}$ = 2.35$\pm$0.04 makes
this source the faintest ``detection'' in the sample, F1042M =
19.18$\pm$0.08, very close to the detection limits of the image.
No counterpart is seen in the F814W image, suggesting a very red
source, F814W--F1042M $>$ 6.2.

However, it is quite likely that
this object is simply a residual cosmic ray.  Figure 7 shows the
two original PC exposures of the 2MASS 1217$-$0311 field, along with the
final combined image, a single PSF-subtracted image,
and a double PSF-subtracted image.
Cosmic rays show up as bright pedestals of flux,
in contrast to the more gradually sloping PSF.
The potential companion (to the right of the primary target,
indicated by an arrow in the fourth panel of Figure 7)
is within one pixel of a bright, extended cosmic
ray in the first image, and is completely wiped out by a cosmic ray
in the second image.  It is
likely that the source in the first
image is a cosmic ray itself.  Only 3--4 overlapping (i.e., detected on
the same pixels in both exposures) cosmic rays were found
within 25 pixels of any of the target images, most of which were quite
obviously identifiable.  This source is less obvious given its
more gradually sloped profile (i.e., not a flat pedestal)
and very faint flux.  Fitting two PSFs to the cosmic ray-corrected
image results in significantly
reduced residuals (by roughly 10\%), although the fainter source is not completely subtracted
out as a result of this fit (fifth panel in Figure 7).  Such a fit
is inconclusive, however, as subtracting any PSF from a cosmic ray will
reduce the overall image residuals, while faint legitimate sources may not subtract
cleanly due to the increased relative noise.
We therefore classify this source
as a candidate companion, requiring additional followup to verify its
existence.

If the companion were real, it would be an extremely interesting object,
as its color limit and intrinsic faintness would make it the
coldest and faintest brown dwarf so far identified.  Assuming
simplistically that $\log{T_{eff}}$ scales with F1042M magnitude,
and using the absolute magnitudes and estimated T$_{eff}$s of
Gliese 229B \citep[$\sim$950 K]{mar96} and Gliese 570D
\citep[$\sim$800 K]{me00a,geb01}, we estimate T$_{eff}^B$/T$_{eff}^A$
$\approx$ 0.6 and hence T$_{eff}^B$ $\approx$ 500 K.  Because
H$_2$O begins to condense in atmospheres as cool as this
\citep{bur99}, 2MASS 1217$-$0311B would probably not be a T dwarf
but the prototype for a new spectral class.  It would also have an
extremely low mass, M $\approx$ 0.012 M$_{\sun}$ for an age of 1
Gyr \citep{bur97}. Hence, confirmation of this possible companion
by follow-up imaging is clearly a priority.

\subsection{Single Sources}

Two other T dwarfs in our sample warrant additional discussion.
The first is the bright (J = 13.83$\pm$0.03) T dwarf
2MASS 0559$-$1404 \citep{me00c}.  This object has a measured parallax
\citep{dah02}, and
is over 1 mag brighter at J-band than the L8 dwarfs 2MASS
1632+1904 \citep{kir99} and Gliese 584C \citep{kir00}, and only
0.6 mag fainter at K$_s$.  \citet{me01} has suggested that 2MASS
0559$-$1404 may be an equal-magnitude binary based on its
brightness and derived T$_{eff}$, although an alternate hypothesis
may be the rapid dissolution of dust cloud material over the L/T
transition \citep{me02b}. Our images rule out the presence of
bright secondary closer than 0$\farcs$05. If this hypothetical
companion exists, it is either currently aligned with the primary
or is separated by less than 0.5 AU. As at least one brown dwarf
spectroscopic binary has been found with a separation less than
this limit \citep[PPl 15]{bas99}, high-resolution spectroscopy of
2MASS 0559$-$1404 may be required to fully rule out the presence
of a close companion.

2MASS 1237+6526 is another T dwarf whose duplicity is under
consideration.  \citet{me00b} have suggested that the unique
H$\alpha$ emission in this object may be due to the presence of a
close ($a$ $\lesssim$ 0.003 AU), interacting companion, although
\citet{me02c} have failed to find photometric evidence of an
eclipsing system.  Our HST images do not rule out the presence of
this hypothetical companion, as a spatial resolution of
0$\farcs$0002 (assuming a distance of 14 pc) would be required to
resolve it.  The very red F814W$-$F1042M and F814W$-$J colors of
this T6.5 dwarf, similar to the T8 Gliese 570D, could arise from
warm circum(sub)stellar material, consistent with this object
being a very young and low-mass (3--12 M$_{Jup}$) weak-line T
Tauri object \citep{me02c,lie02}.  Photometry from \citet{dah02} confirm
this object's very red optical/near-infrared colors, but give no
evidence for a reddened J$-$K color ($-$0.26$\pm$0.20). Hence, the
nature of this object remains ambiguous.

\section{Binary Statistics}

\subsection{The Binary Fraction}

Of the ten T dwarfs imaged in our sample, two have clearly resolved
binary companions, implying an observed binary fraction of
20$^{+17}_{-7}$\%\footnote{Derivation of uncertainties for all
sample statistics are described in the Appendix.}. This is
consistent with results obtained for field late-type M and L
dwarfs \citep{koe99,rei01a,clo02}.

One must keep in mind, however, that the T dwarf sample was
originally drawn from a magnitude-limited search, and that the
observed binary fraction (for initially unresolved pairs) is
therefore biased \citep{opi24,bra76}.  If we assume negligible
contribution by multiple systems of three components or more, then
the observed binary fraction, ${\epsilon}_b^{obs} \equiv
N_{binary}^{obs}/N_{total}^{obs}$, is related to the ``true'', or
volume-limited, fraction, ${\epsilon}_b \equiv
N_{binary}/N_{total}$, by
\begin{equation}
{\epsilon}_b^{obs} = {\alpha}\frac{{\epsilon}_b}{1-{\epsilon}_b+{\alpha}{\epsilon}_b},
\end{equation}
where
\begin{equation}
{\alpha} \equiv \frac{\int_0^1{(1+{\rho})^{3/2}f({\rho})d{\rho}}}{\int_0^1{f({\rho})d{\rho}}}
\end{equation}
is the fractional increase in volume sampled for binaries with
flux ratio ${\rho} \equiv F_B/F_A$ and flux ratio distribution
$f({\rho})$. In the case of all binaries being equal-magnitude
systems, $\alpha$ = 2$^{3/2}$ = 2.8; while in the case of a flat
$f({\rho})$, $\alpha$ = 1.9. From these limiting cases and our
observed binary fraction, we derive ${\epsilon}_b =
9^{+15}_{-4}$\%, where we have included the uncertainty in
${\epsilon}_b^{obs}$ and allowed $\alpha$ to vary between 1.9 and 2.8.
This value is significantly
lower than the binary fraction of more massive stars \citep{duq91,fis92,rei97}.

A second means of obtaining a bias-free estimate of the binary
fraction in our magnitude-limited sample is by computing the
relative luminosity functions with the $1/V_{max}$ technique
\citep{scm68,scm75}.  Simply, ${\epsilon}_b =
{\Phi}_{binary}/{\Phi}_{total}$, where is $\Phi$ is the luminosity
function calculated from
\begin{equation}
{\Phi} = \sum_{i=1}^N{\frac{1}{V^{(i)}_{max}}},
\end{equation}
the sum carried over all N objects in the sample, with $V_{max}
\propto d^3_{max} - d^3_{min}$.  For the T dwarfs, the minimum
detectable distance, $d_{min}$, is set by the constraint of no
optical counterpart in the USNO A2.0 catalog \citep{mon98} or in
Digital Sky Survey images, and is roughly 1 pc for all objects in
our sample \citep{me01}; the maximum detectable distance,
$d_{max}$, depends on the sample search limit (J = 16) and the
absolute J magnitude of the object. We have estimated $M_J$ for
objects in our sample using the known
absolute magnitudes of the T5 2MASS 0559$-$1404, T6 SDSS 1624+0029,
T6.5 Gliese
229B, and T8 Gliese 570D, interpolating by spectral type
\citep{me01}. Because binaries can be detected to
distances $\sqrt{1+{\rho}}$ further than single objects, we have
included this correction for 2MASS 1225$-$2739AB and 2MASS
1534$-$2952AB using ${\rho}_J$ = 0.29 and 1.0, respectively.  The
derived binary fraction is only 6\%, on the low end of, but not
inconsistent with, the bias-corrected value given above, and again
much lower than stellar binary fraction measures. An
estimate of uncertainty for this technique is not straightforward
\citep{men01}; nonetheless, the value is consistent with a binary
fraction much less than that of more massive stellar systems.  The
completeness estimator for our sample \citep{scm68}, $<V/V_{max}>$
= 0.51$\pm$0.09, gives us some confidence that our result is not
significantly influenced by incompleteness or color bias.

The bias correction given in Eqn.\ 4 is applicable to the L dwarf
sample of \citet{rei01a}, which is also based on a
magnitude-limited survey of the 2MASS database
\citep{kir99,kir00}.  They found ${\epsilon}_b^{obs}$ =
20$^{+12}_{-6}$\% (4 of 20), which translates into a corrected
fraction ${\epsilon}_b = 9^{+11}_{-4}$\%, where again we have
included uncertainty in both ${\epsilon}_b^{obs}$ and $\alpha$.
The 1/$V_{max}$ technique gives a consistent value of 12\%; again,
significantly lower than the binary fraction of more massive
stars.

Hence, we find that both L and T dwarf samples, when corrected for
selection bias, yield binary fractions which are significantly
lower than measurements made for more massive stars, suggesting a
continuation of the apparent trend of decreasing ${\epsilon}_b$ from F--G
to M dwarf stars. However, it must be stressed that the derived
fraction is applicable for separations $a \gtrsim 1-5$ AU and $q
\gtrsim 0.4$, while the investigation of, e.g., \citet{duq91},
probed much smaller mass ratios ($q \rightarrow 0.1$) and
separations ($a \rightarrow 0.1$ AU). Hence, our sample may
contain binaries with secondaries below our detection limits, or
very tight unresolved binaries. We can estimate the contribution
from the latter
population by examining the frequency of M dwarf
spectroscopic binaries in the
magnitude-limited sample of \citet{rei02}, who found 6$^{+6}_{-2}$\%
(2 of 36) of their targets were spectroscopic
binaries. Again, using Eqn.\ 4, this implies a bias-corrected
fraction of only 3$^{+3}_{-2}$\%, increasing the net binary
fraction of the T dwarfs
to perhaps 12\%, not enough to bring our results in
agreement with the binary fraction of F--G or M dwarfs.
Similarly, if we compare our derived binary frequency to only
those stellar systems having $a \gtrsim 5$ AU, roughly 41\% in the
\citet{duq91} F--G dwarf sample (using their Gaussian log(P) distribution) and
31\% in the \citet{fis92} M dwarf sample (their Table 2),
we find that there are clearly
fewer T dwarf multiple systems in this separation regime.  The
contribution of lower-mass companions is also insufficient to
explain the deficiency of L and T dwarf binaries,
as binary fractions for F--G and M dwarf systems with 0.4 $< q <$ 1.0
are roughly 33\% (from Table 7 in Duquennoy and Mayor 1991)
and 32\% (from Fischer \& Marcy 1992, assuming a flat mass ratio distribution),
respectively, significantly higher than our results.
Hence, unless T dwarfs prefer very closely-separated (see below)
and/or very low-mass companions, the binary fraction of these
objects is significantly lower than that of more massive stars.

\subsection{Separation Distribution}

The two confirmed binary systems identified in this survey have
projected separations $a$ $\lesssim$ 3 AU, and no wide, co-moving
companions to any of these objects have yet been identified in
either the HST data or the 2MASS survey. In fact, no wider
companions ($a$ $>$ 2$\arcsec$) have been found around any T dwarf
identified in the 2MASS or SDSS surveys. This result is consistent
with the current absence of widely-separated late-M and L dwarf
binaries (Table 7), all of which have $a$ $\lesssim$ 10--20 AU. In
contrast, roughly 50\% of the more massive early-type M dwarf multiple
systems in the \citet{fis92} study have 10 AU $\lesssim$ $a$
$\lesssim$ 10$^4$ AU. Similarly, roughly 40\% of M dwarf multiple
systems in the 8 pc sample have separations greater than 10 AU
\citep{rei97}. If lower-mass systems had a similar fraction of
wide binaries, then roughly 20 pairs with $a > 10-20$ AU from the
approximately 300 known L and T dwarfs should have been
identified, while there are currently none.  The absence of wide systems
may contribute to the overall deficiency in multiple systems
amongst the T dwarfs, as the binary fractions of
F--G dwarfs and M dwarfs drop to roughly 20\% for
separations $a < 10$ AU, within the uncertainty estimates of our
bias-corrected result.  We discuss the apparent limit in the
separations of low-mass stars and brown dwarfs further in $\S$7.1.

\subsection{Mass Ratio Distribution}

Finally, we consider the mass ratio distribution, $f(q)$, a
statistic that can constrain the origin of secondaries in a binary
population. In general, masses are difficult to derive for field
brown dwarfs, as estimates depend on both temperature and age, and
there are few empirical clues currently known
for the latter parameter. In Table
7, we have estimated masses for 2MASS 0746+2000AB, DENIS
1228$-$1159AB, 2MASS 1534$-$2952AB, and 2MASS 1225$-$2739AB
assuming an age of 1 Gyr, the T$_{eff}$ scale of \citet{me02}, and
the theoretical models of \citet{bur97}; for 2MASS 1146+2230AB, we
used maximum masses of 0.06 M$_{\sun}$ based on the presence of
the 6708 {\AA} Li I line in the combined light spectrum
\citep{kir99}. All other mass estimates are taken from the listed
references. Fortunately, the desired quantity, $q$, is not greatly
sensitive to these assumptions.

The two T dwarf binaries identified in our sample have relatively
large mass ratios, $q$ = 0.8 and 1.0.  As discussed
in $\S$4.3, we were capable of identifying systems down to $q =
0.4$, albeit not for very closely separated systems like 2MASS 1534$-$2952AB.
When we place these two systems in context with the other
low-mass binaries listed in Table 7, there appears to be a
preference for
equal-mass systems, the lowest mass ratio system being 0.7. This is
similar to what has been reported in the 8 pc sample
\citep{rei97}, and is at odds with the flatter distributions of
\citet{duq91} and \citet{fis92}.

The two binaries identified in our program form too small a sample
to examine the mass ratio distribution statistically, so we
combined our ten T dwarfs with the L dwarf sample of
\citet{rei01a}. Based on their completeness limits, and using the
same mass/flux ratio scaling as described in $\S$4.3, we find that
their sample is complete to $q \gtrsim 0.4$ for $a \gtrsim
0{\farcs}25$, or $a \gtrsim 5$ AU assuming a mean distance of 20
pc. This is comparable to our completeness for $a \gtrsim 4$ AU,
although the inner separation limit for our sample is roughly one-half
that of the L dwarf study.  Nonetheless, because similar instruments
and observing strategies were employed, combining these two
samples should not introduce significant biases.  The observed
binary fraction for this combined sample is 20$^{+9}_{-5}$\%,
while the bias-corrected fraction is $9^{+9}_{-3}$\%. Breaking
these systems down by mass ratio, we find 4$^{+0.8}_{-1.3}$
systems with $1.0 \leq q < 0.9$, 1$^{+1.4}_{-0.4}$ system with
$0.9 \leq q < 0.8$, 1$^{+1.4}_{-0.4}$ system with $0.8 \leq q <
0.7$, and less than 1.4 systems for all other ratio bins.  We plot
this distribution (light grey histogram) in Figure 8, normalized
so that $f(q=1) = 1$.

Again, there appears to be a preference for equal mass binaries.
This is not unexpected, however, given the intrinsic faintness of
very low-mass brown dwarfs, and the preferential selection of
equal-mass systems in magnitude-limited surveys.  We must,
therefore, consider selection biases in these magnitude-limited
samples. The correction to the flux ratio distribution is
\begin{equation}
\frac{f({\rho})^{obs}}{f({\rho})} \propto (1+{\rho})^{3/2},
\end{equation}
which, using the mass/flux ratio scaling as before, yields
\begin{equation}
\frac{f({q})^{obs}}{f({q})} \propto (1+q^{2.64})^{3/2}.
\end{equation}
Hence, the bias is a fairly strong function of $q$.  Applying
these corrections, we derive a slightly revised mass ratio
distribution (dark grey histogram in Figure 8). Even with the bias
corrections, there are more objects in the $0.9 \leq q < 1.0$ bin
than in other mass ratio bins, although not a statistically
significant number.  With the substantial
statistical uncertainties of our small sample, in particular the
large upper limits for $q \lesssim 0.6$, we cannot statistically
rule out a flatter distribution.

We have also plotted the M dwarf $f(q)$ for the 8 pc sample
\citep{rei97} in Figure 8, which we have normalized as above and computed
uncertainties based on a total of 21 M dwarf multiple systems.
This distribution appears to be similar to the L and T dwarf
distribution, with a possible preference for high-mass ratio
systems, although the uncertainties are again significant;
a flatter
distribution cannot be statistically ruled out.
Hence, concluding a preference of equal-mass components amongst the
M, L, and T dwarf binaries requires considerably better
statistics, but our results are suggestive of this trend.

\section{Brown Dwarf Binary Formation and Disruption}

The results above indicate that both the binary fraction and
separation distribution of brown dwarfs are significantly
different that those of more massive stars, while the mass ratio
distribution suggests a preference for equal-mass systems. We now
examine how these properties may constrain the formation or
evolution of substellar binary systems.

\subsection{Disruption by Stellar and GMC Encounters}

The deficiency of brown dwarf binaries with $a \gtrsim 10$ AU is
reminiscent of the deficiency of stellar binaries with $a$
$\gtrsim$ 0.1 pc $\approx$ 2$\times$10$^5$ AU
\citep{bah81,clo90,was91}. While there remains some debate as to
whether a sharp break exists in the separation distribution
\citep{ret82,was87,was91,clo90}, it is generally believed that impulsive
perturbations by close stellar encounters or passage through a GMC
causes a gradual diffusion of separations and binding energies,
ultimately resulting in the dissolution of weakly bound systems in
a catastrophic encounter \citep{wei87}. Because the binding
energies of brown dwarf pairs are small and such systems therefore
easily disrupted, it is tempting to ascribe the same mechanism to
the apparent absence of widely-separated systems.

To examine the probability of disruption by stellar and GMC
encounters, we used the formalism of \citet{wei87}, adopting the
general parameters used by the authors\footnote{V$_{rel}$ = 20 km
s$^{-1}$, $\epsilon$ = 0.1, $n_*$ = 0.05 pc$^{-3}$, $n_{GMC}$ =
4$\times$10$^{-8}$ pc$^{-3}$, R$_{GMC}$ = 20 pc, M$_{GMC}$ =
5$\times$10$^5$ M$_{\sun}$, and N$_{clump}$ = 25; see
\citet{wei87} for nomenclature.} and examine the evolution of two
0.05 M$_{\sun}$ gravitationally bound brown dwarfs with separation
10 AU. The critical impact parameter for significant gravitational
disruptive effects is $b_{max} \propto a^{3/2}M^{-1/2} \approx 70$
AU. For stellar encounters, the Fokker-Planck impact parameter in
the tidal limit\footnote{In the case we are considering,
$GM/{\epsilon}aV_{rel}^2 > (M/M_*) \sim 0.1$, so that, unlike wide
stellar pairs, the tidal limit applies.}, $b^*_{FP} \propto
a^{3/4}M^{-1/4} \approx 30$ AU $\approx a$, implies that both
close, catastrophic collisions and gradual tidal disruption can
affect the evolution of brown dwarf pairs. However, the frequency
of close stellar encounters is ${\Gamma}^*_{cat} \propto aM^{-1}
\approx (2{\times}10^5$ Gyr)$^{-1}$, while the diffusive timescale
is ${\tau}^* \propto a^{-1}M \approx 3700$ Gyr. Hence, stellar
encounters are not frequent enough to affect brown dwarf binaries
with $a \lesssim 10^4$ AU over the age of the Galaxy.  The
tidal limit impact parameter for GMC interactions is $b^{GMC}_{FP}
\approx 2{\times}10^5$ AU $>> b_{max}$, while the impact parameter
for catastrophic interactions with GMC clumps in the case of cloud
penetration (occurring at a rate of roughly 1 Gyr$^{-1}$) is
$b^{clump}_{FP} \approx 2000$ AU $>> b_{max}$; hence, GMC
interactions play no role in the disruption of brown dwarf binary
systems. Therefore, the separation limit of brown dwarf pairs is
not due to disruption in the Galactic field, as appears to be the
case for wide stellar binaries.  Indeed, only brown dwarf systems
with separations many orders of magnitude larger than those
observed could be disrupted in the field.

To further elucidate how stellar and GMC disruptions do not
constrain the separation of brown dwarf pairs, Figure 9 plots the
separation of binary stars and brown dwarfs versus total mass.
Binary data for brown dwarf and late-type stars (primaries later
than M8) are listed in Table 7; for stellar binaries, we include
the samples of \citet{clo90}, \citet{duq91}, \citet{fis92},
\citet[HST M dwarf binaries and the 8 pc sample]{rei97}, and
\citet[Multiple Star Catalog]{tkv97}; finally, for stellar-brown
dwarf binaries we use compiled values from \citet{rei01a}. The
absence of wide low-mass pairs is quite striking in this figure,
particularly given the ability of 2MASS and other field surveys to
resolve such systems.  The curved line shows a log-normal relation
for the maximum separation of binary systems, $\log{a_{max}} =
3.33M_{tot}+1.1$, which is appropriate for disruption by
point-source encounters \citep{rei01a}. Note that such an envelope
matches the observed cut-off for stellar binaries quite well, but
allows more-widely separated brown dwarf binaries to form ($a
\approx 20-30$ AU).  For the lower-mass systems, we find a second
line,
\begin{equation}
a_{max}(AU) = 1400{\times}M_{tot}^2,
\end{equation}
appears to be more adequate for the separation limit. While the
number of objects for which this envelope applies is relatively
small, it is not biased by selection effects, as all systems would
be unresolved in their original surveys, and would be easily
resolved by HST, for $a > a_{max}$.  We suggest that this
power-law relation may be a clue to the mechanism that modulates
the formation or disruption of substellar binaries, although further
data are required to confirm if this relation is truly representative
of all brown dwarf binary systems.

What about disruption within the nascent star-forming cluster?  A
survey of the $\sim$ 120 Myr Pleiades cluster by \citet{mrt00b}
found no binaries out of a sample of 34 low-mass star or brown
dwarf members for $a \gtrsim 27$ AU, although candidate
photometric binaries (including PPL 15) suggest a binary fraction
of $\sim 22$\% \citep{rei01a}, consistent with the binary fraction
observed in the field.
\citet{duc99} found no substellar binaries or companions in the
0.5--10 Myr IC 348 cluster for $a$ $\gtrsim$ 30 AU, although one
very wide ($a$ $\approx$ 2300 AU) candidate system in this cluster
has been suggested by \citet{naj00}. Finally, a search for binary
objects in 1--5 Myr Cha H$\alpha$ 1 cloud by \citet{neh02} and
\citet{neh03} has turned up only one potential binary candidate
with $a$ $\lesssim$ 28 AU. Therefore, it appears that for
disruption to play an important role in the elimination of brown
dwarf binaries with $a \gtrsim 10$ AU, it must occur within a few
million years of formation.  Note that the theory of \citet{wei87}
predicts that stellar encounters may have some influence in young
dense clusters, as ${\Gamma}^*_{cat} \propto n_* \sim 1$
Gyr$^{-1}$, and ${\tau}^* \propto n^{-1} \sim 20$ Myr for $n \sim
10^4$ pc$^{-3}$, typical for regions such as the Orion Nebular
Cloud \citep{hil97}. However, as these dense regions rapidly
disperse (i.e., within a few Myr), close encounters are probably
not solely responsible for the absence of widely-separated
substellar binaries.

\subsection{Small N Protoclusters and Brown Dwarf Ejection Models}

A currently popular model of star formation in clusters is through
the fragmentation of molecular clouds into small aggregates of
non-hierarchical protostellar cores \citep{lar72}, with the entire
young star forming region being comprised of these initial
groupings. On a short timescale (ages $\lesssim 10^5$ yr), these
``protoclusters'' are disrupted by dynamical interactions between
the cores, which is likely modulated by residual gas and dust that
continues to be accreted \citep{bon97,bon01}.  Such dynamical
interactions preferentially eject the lowest-mass components,
while an ejected core is also less likely to continue significant
accretion.  These considerations have given rise to so-called
``ejection'' formation models for brown dwarfs
\citep{rpt01,bat02}, in which the dynamic removal of cores from
accretion regions condemns them to remain below the Hydrogen
burning minimum mass.  A numerical simulation by \citet{bat02}
utilizing this general model have found a low brown dwarf binary
fraction, at most 5\%, based on a single remaining undisturbed
pair in a dynamically unstable multiple system.  This fraction is
consistent with the derived fraction of L and T dwarf binaries,
although the simulation also predicts similarly low binary fractions for
low-mass stars, which is not observed.
Nonetheless, since the ejection model predicts the
disruption of potential brown dwarf binaries at very early ages,
while also imposing a limit to the dimensions of such systems
\citep{rpt01}, it shows some promise in explaining the origins of
substellar systems in general.

\subsection{Fragmentation}

The preference for brown dwarfs to form close binaries may not
necessarily require a disruptive process, however.  Studies of
young binary stars favor fragmentation \citep{bos88} as the
dominant mode of binary formation, due to coevality of components,
the presence of circumbinary structures, and the preference for
equal-mass components in closely-separated systems \citep{whi01}.
These conditions do not require dynamical disruption from
neighboring protostellar systems.  In general, a low-mass gas and
dust core must collapse to smaller dimensions before it achieves
sufficient densities to continue fragmentation, producing multiple
systems which are initially closely separated.  This suggests a
maximum separation dependence on mass, as hinted at in Figure 8,
although no theoretical prediction as such has been made.
The deficiency of low-mass pairs may arise from the inability for
very small cloud clumps to both form and also continue
fragmenting, although the influence of magnetic fields,
turbulence, and external perturbations would also have substantial
influence. Current models (e.g., Boss 2001) are capable of
producing core fragments in the range of 10s of Jupiter masses
(M$_{Jup}$), in the mass range of field L and T brown dwarfs,
but masses down to 1 M$_{Jup}$ require dynamical
ejection to prevent further accretion.

\subsection{How Do Brown Dwarfs Form?}

The similarity in the binary fractions and
separation distributions for young cluster and field low-mass
systems, and the low probability of dynamic disruption in all but
the densest stellar environments, makes it highly probable that the
field brown dwarf binary distribution is quite similar to the
natal distribution.  This is important, as the distances and dust
opacity of protostellar environments, and the relative faintness
of protosubstellar objects, makes investigation of brown dwarf
formation at very early ages quite difficult.
Improving the statistics for field brown dwarf systems, and examining closer
separation regimes through radial velocity techniques, should
provide considerable insight into the formation of these very
low-mass objects.  We find that
both fragmentation and ejection models produce some of the
qualitative characteristics of late-M, L, and T dwarf binaries, and
it is possible that substellar systems form by some combination of
these processes.  However, more detailed quantitative predictions must
be matched with large, unbiased sample statistics before conclusive
statements can be made on the formation of brown dwarfs.

\acknowledgments

We thank our referee, L.\ Close, for in-depth criticisms and helpful
suggestions for our manuscript, and useful discussions on wide stellar
binaries.  We also thank A.\ Ghez, D.\ Koerner, \& J.\ Liebert for discussions on
disks and binary star formation; and D.\ Koerner and A.\ Dolphin
for useful discussions on PSF fitting. AJB acknowledges support by
NASA through Hubble Fellowship grant HST-HF-01137.01 awarded by
the Space Telescope Science Institute, which is operated by the
Association of Universities for Research in Astronomy, Inc., for
NASA, under contract NAS 5-26555. JDK acknowledges the support of
the Jet Propulsion Laboratory, California Institute of Technology,
which is operated under contract with the National Aeronautics and
Space Administration. Based in part on observations made with the
NASA/ESA Hubble Space Telescope, obtained at the Space Telescope
Science Institute, which is operated by the Association of
Universities for Research in Astronomy, Inc., under NASA contract
NAS 5-26555. These observations are associated with proposal ID
8563. This publication makes use of data from the Two Micron All
Sky Survey, which is a joint project of the University of
Massachusetts and the Infrared Processing and Analysis Center,
funded by the National Aeronautics and Space Administration and
the National Science Foundation.

\appendix

\section{Probability Distribution for the Binary Fraction}

When binary fractions (or other equivalent frequency statistics)
are quoted in the literature, they are frequently assigned Poisson
uncertainties. However, the Poisson limit applies only in the case
of a large sample, whereas the brown dwarf samples discussed here
are less than 30 in number.  Hence, we derived statistical
uncertainties by constructing a probability distribution for
${\epsilon}_b$ given the total sample size, $N$, and the number of
binaries in the sample, $n$.  The binomial distribution determines
the probability of finding $n$ binaries given the sample size and
binary fraction, as:
\begin{equation}
B(n;N,{\epsilon}_b) =
\frac{N!}{n!(N-n)!}{\epsilon}_b^n(1-{\epsilon}_b)^{N-n}.
\end{equation}
However, this equation may also be used to derive the probability
distribution of ${\epsilon}_b$ given the observed quantities N and n.
To do this, we compute
$B^{\prime}({\epsilon}_b;n,N) \propto B(n;N,{\epsilon}_b)$ for $0
\leq {\epsilon}_b \leq 1$, normalizing
\begin{equation}
\int^1_0{B^{\prime}({\epsilon}_b;n,N){\rm d}{\epsilon}_b} = 1,
\end{equation}
which yields $B^{\prime} = (N+1)B$.

Figure 9 plots $B^{\prime}$ for our T dwarf sample, $N = 10$ and
$n = 2$.  To derive upper and lower uncertainty limits, ${\epsilon}^U_b$ and
${\epsilon}^L_b$, we computed the values for
which $\int_0^{{\epsilon}^U_b}{B^{\prime}{\rm d}{\epsilon}_b}$ =
$\int_{{\epsilon}^L_b}^{1}{B^{\prime}{\rm d}{\epsilon}_b}$ = 0.84,
equivalent to 1$\sigma$ limits for a Gaussian distribution.
These limits can also be found numerically by solving
\begin{equation}
\sum_{i=0}^n{\frac{(N+1)!}{i!(N+1-i)!}x^i(1-x)^{N+1-i}} = \left\{ \begin{array}{l}
 0.84, x = {\epsilon}^L_b \\
 0.16, x = {\epsilon}^U_b \\
 \end{array}
 \right. .
\end{equation}
As shown in Figure 9, the derived limits are not
symmetric about the probability peak, prohibiting ranges which
exceed the sample size or are less than zero.  For large samples
($N \gtrsim$ 100) one recovers the standard Poisson uncertainty
limits, (${\epsilon}^U_b-{\epsilon}_b)/{\epsilon}_b =
({\epsilon}_b-{\epsilon}^L_b)/{\epsilon}_b = \sqrt{1/n+1/N}$.

\clearpage

\begin{deluxetable}{lllcclcc}
\tabletypesize{\scriptsize}
\tablecaption{Log of HST Observations.}
\tablewidth{0pt}
\tablehead{
 & & \multicolumn{2}{c}{F814W} & & \multicolumn{2}{c}{F1042M} & \\
\cline{3-4} \cline{6-7} \colhead{Object\tablenotemark{a}} &
\colhead{SpT} & \colhead{UT Date/Time\tablenotemark{b}} &
\colhead{$t$ (sec)} & & \colhead{UT Date/Time\tablenotemark{b}} &
\colhead{$t$ (sec)} &
\colhead{RA ($\degr$)\tablenotemark{c}} \\
\colhead{(1)} & \colhead{(2)} & \colhead{(3)} & \colhead{(4)}  & &
\colhead{(5)} & \colhead{(6)} & \colhead{(7)}  }
\startdata
2MASS 05591914$-$1404488 & T5   & 20000906 18:41 & 2400  &  & 20000906 20:15  & 2600  &  305 \\
2MASS 09373487+2931409 &  T6pec  & 20001016 23:44 & 2600  & & 20001017 00:08 & 2600  &  339 \\
2MASS 10475385+2124234 &  T6.5   & 20010104 16:16  & 2400  & & 20010104 17:50 & 2600  &  310 \\
2MASS 12171110$-$0311131 & T7.5   &  20000704 03:10   & 2400 & & 20000704 04:43 & 2600  & 158  \\
2MASS 12255432$-$2739466 & T6   &  20010410 15:53   & 2600  &  & 20010410 17:29 & 2600  & 79 \\
2MASS 12373919+6526148 &  T6.5  & 20000613 15:32  & 2000  & & 20000613 16:49  & 2400  & 160 \\
Gliese 570D &  T8   & 20000818 07:23 & 2400    &  & 20000818 08:54 & 2600  &  152 \\
2MASS 15344984$-$2952274 &  T5.5   & 20000818 04:10  & 2600  & & 20000818 05:42  & 2600  & 148 \\
2MASS 15462718$-$3325111 &  T5.5   & 20000819 05:55  & 2600  &  & 20000818 07:26  & 2600  & 147 \\
2MASS 23565477$-$1553111 &  T6   & 20001129 09:48  & 2400  & & 20001129 11:23  & 2600  & 66 \\
\enddata
\tablenotetext{a}{Source designations for the 2MASS Point Source Catalog are
given as ``2MASS Jhhmmss[.]ss$\pm$ddmmss[.]s''.  The suffix conforms
to IAU nomenclature convention and is the sexigesimal R.A. and
decl. at J2000 equinox.} \tablenotetext{b}{UT date/time given as
yyyymmdd hh:mm.} \tablenotetext{c}{Telescope Roll Angle East
from North.}
\end{deluxetable}

\begin{deluxetable}{cccccc}
\tabletypesize{\scriptsize} \tablecaption{WPFC2 Aperture
Corrections. \label{tab:apcor}} \tablewidth{0pt} \tablehead{
\colhead{Filter} & \colhead{Aperture\tablenotemark{a}} &
\colhead{PC} & \colhead{WF1} & \colhead{WF2} &
\colhead{WF3} \\
\colhead{(1)} &
\colhead{(2)} &
\colhead{(3)} &
\colhead{(4)} &
\colhead{(5)} &
\colhead{(6)}
}
\startdata
 & N$_{*}$ & 24 & 66 & 79 & 78 \\
 & 2 & 1.62$\pm$0.07 & 1.21$\pm$0.03 & 1.21$\pm$0.02 & 1.21$\pm$0.02 \\
F814W & 3 & 1.21$\pm$0.04 & 1.09$\pm$0.02 & 1.08$\pm$0.01 & 1.08$\pm$0.01 \\
 & 5 & 1.04$\pm$0.02 & 1.02$\pm$0.02 & 1.01$\pm$0.01 & 1.01$\pm$0.01 \\
 &  &  &  &  &  \\
 & N$_{*}$ & 7 & 21 & 17 & 19 \\
 & 2 & 1.65$\pm$0.08 & 1.28$\pm$0.06 & 1.31$\pm$0.05 & 1.28$\pm$0.05 \\
F1042M & 3 & 1.33$\pm$0.09 & 1.11$\pm$0.04 & 1.13$\pm$0.03 & 1.12$\pm$0.03 \\
 & 5 & 1.01$\pm$0.06 & 1.03$\pm$0.03 & 1.03$\pm$0.02 & 1.03$\pm$0.02 \\
\enddata
\tablenotetext{a}{Aperture in pixels, corresponding to angular apertures
of 0$\farcs$09 (0$\farcs$20), 0$\farcs$14 (0$\farcs$30), and 0$\farcs$23 (0$\farcs$50)
for 2, 3, and 5 pixels on the PC (WF) chips.}
\end{deluxetable}

\begin{deluxetable}{llcccccc}
\tabletypesize{\scriptsize}
\tablecaption{T Dwarf Photometry.}
\tablewidth{0pt}
\tablehead{
\colhead{Object} &
\colhead{SpT} &
\colhead{F814W} &
\colhead{F1042M} &
\colhead{2MASS J} &
\colhead{F814W$-$F1042M} &
\colhead{F1042M$-$J} &
\colhead{F814W$-$J} \\
\colhead{(1)} &
\colhead{(2)} &
\colhead{(3)} &
\colhead{(4)} &
\colhead{(5)} &
\colhead{(6)} &
\colhead{(7)} &
\colhead{(8)}
}
\startdata
2MASS 0559$-$1404 & T5 & 18.65$\pm$0.03 &  15.02$\pm$0.07 &   13.83$\pm$0.03 &   3.64$\pm$0.08 & 1.19$\pm$0.08 & 4.82$\pm$0.04 \\
2MASS 1534$-$2952AB\tablenotemark{a} & T5.5  & 19.62$\pm$0.02 & 15.75$\pm$0.07 &   14.90$\pm$0.04    & 3.86$\pm$0.08 & 0.85$\pm$0.08 & 4.72$\pm$0.04 \\
2MASS 1546$-$3325 & T5.5  & 20.52$\pm$0.03 &  16.66$\pm$0.07 &   15.60$\pm$0.05    & 3.86$\pm$0.08 & 1.06$\pm$0.09 & 4.92$\pm$0.06 \\
2MASS 1225$-$2739A\tablenotemark{b} & T6  & 20.32$\pm$0.03 &  16.38$\pm$0.07 &   15.50$\pm$0.05    &  3.94$\pm$0.08 & 0.88$\pm$0.09 & 4.83$\pm$0.06 \\
2MASS 2356$-$1553 & T6  & 20.73$\pm$0.03 &  16.96$\pm$0.08 &   15.80$\pm$0.06  &   3.77$\pm$0.08 & 1.16$\pm$0.10 & 4.93$\pm$0.07 \\
2MASS 0937+2931 & T6pec  & 19.73$\pm$0.03 &  15.47$\pm$0.07 &   14.65$\pm$0.04  &  4.26$\pm$0.08 & 0.82$\pm$0.08 & 5.08$\pm$0.05 \\
Gliese 229B\tablenotemark{c} & T6.5  & 19.49$\pm$0.03  & 15.16$\pm$0.03  &     14.32$\pm$0.05 &  4.33$\pm$0.04 & 1.15$\pm$0.06 & 5.17$\pm$0.06  \\
2MASS 1047+2124 & T6.5  & 20.89$\pm$0.03 &  16.67$\pm$0.07 &   15.82$\pm$0.06    & 4.22$\pm$0.08 &  0.85$\pm$0.10 &  5.07$\pm$0.07 \\
2MASS 1237+6526 & T6.5  & 21.25$\pm$0.03 &  16.76$\pm$0.07 &   16.03$\pm$0.09    & 4.49$\pm$0.08 & 0.73$\pm$0.12 & 5.22$\pm$0.10 \\
2MASS 1217$-$0311 & T7.5  & 21.11$\pm$0.03 &  16.83$\pm$0.07 &    15.85$\pm$0.07 &   4.29$\pm$0.08 &  0.98$\pm$0.10 &  5.27$\pm$0.08 \\
Gliese 570D  & T8  & 20.55$\pm$0.03 &  16.08$\pm$0.07 &   15.33$\pm$0.05 &  4.49$\pm$0.08 &  0.73$\pm$0.09 &  5.22$\pm$0.06 \\
2MASS 1225$-$2739B\tablenotemark{b,d} & T8:\tablenotemark{e}  & 21.91$\pm$0.05 &  17.45$\pm$0.06 &    16.85$\pm$0.08   & 4.46$\pm$0.07 & 0.61$\pm$0.10 & 5.07$\pm$0.09 \\
\enddata
\tablenotetext{a}{WPFC2 magnitudes computed for a 5-pixel
aperture.} \tablenotetext{b}{J magnitude estimated from combined
2MASS  J = 15.22$\pm$0.05 and $\Delta$J = 1.35$\pm$0.08 (see
$\S$3.2).} \tablenotetext{c}{WPFC2 magnitudes from \citet{gol98};
UKIRT J from \citet{leg99}.} \tablenotetext{d}{WPFC2 magnitudes
computed for a 2-pixel aperture.} \tablenotetext{e}{Spectral type
estimated from F814W$-$F1042M color; see $\S$5.1.1.}
\end{deluxetable}

\begin{deluxetable}{lccc}
\tabletypesize{\scriptsize}
\tablecaption{Binary Parameters.}
\tablewidth{0pt}
\tablehead{
 &  \colhead{2MASS 1225$-$2739AB} &
\colhead{2MASS 1534$-$2952AB}  &
\colhead{2MASS 1217$-$0311AB?\tablenotemark{a}} \\
 &  \colhead{(1)} &
\colhead{(2)}  &
\colhead{(3)}
}
\startdata
  SpT & T6/T8 & T5.5/T5.5 & T7.5/Y?  \\
  d (pc)\tablenotemark{b} & 11.2$\pm$0.5 & 16$\pm$5  & 10$\pm$4  \\
 $a$ ($\arcsec$) & 0$\farcs$282$\pm$0$\farcs$005 & 0$\farcs$065$\pm$0$\farcs$007  & 0$\farcs$209$\pm$0$\farcs$006  \\
  $a$ (AU) & 3.17$\pm$0.14 & 1.0$\pm$0.3 & 2.1$\pm$0.8  \\
  $\phi$ ($\degr$) & 250$\pm$7 & 1$\pm$9 & 74$\pm$7   \\
  ${\Delta}$F814W & 1.59$\pm$0.04  &  0.5$\pm$0.3 &  $>$ 4.4   \\
 ${\Delta}$F1042M  & 1.05$\pm$0.03 &  0.2$\pm$0.3 &  2.35$\pm$0.04   \\
\enddata
\tablenotetext{a}{Potential faint companion requiring
confirmation; see $\S$5.1.2.} \tablenotetext{b}{Spectrophotometric
distance estimated from spectral types and T dwarf with known
distances; see $\S$5.1.1.}
\end{deluxetable}

\begin{deluxetable}{lccccccc}
\tabletypesize{\scriptsize}
\tablecaption{Limiting Detection Magnitudes for $a > 0{\farcs}4$.}
\tablewidth{0pt}
\tablehead{
 &  \multicolumn{3}{c}{F814W} & &
\multicolumn{3}{c}{F1042M} \\
 \cline{2-4} \cline{6-8}\noalign{\smallskip}
\colhead{Object} &
\colhead{m$_{lim}$\tablenotemark{a}}  &
\colhead{${\Delta}$M$_{lim}$}  &
\colhead{$q_{lim}$\tablenotemark{b}}  & &
\colhead{m$_{lim}$\tablenotemark{a}}  &
\colhead{${\Delta}$M$_{lim}$}  &
\colhead{$q_{lim}$\tablenotemark{b}} \\
\colhead{(1)} &
\colhead{(2)}  &
\colhead{(3)}  &
\colhead{(4)}  & &
\colhead{(5)}  &
\colhead{(6)}  &
\colhead{(7)}
}
\startdata
2MASS 0559$-$1404 & 25.6 & 6.9 & 0.09 & & 19.9 & 4.9 & 0.18 \\
2MASS 0937+2931 & 25.4 & 5.8 & 0.13 & & 19.9 & 4.4 & 0.22 \\
2MASS 1047+2124 & 25.5 & 4.6 & 0.20 & & 19.9 & 3.2 & 0.33 \\
2MASS 1217$-$0311 & 25.6 & 4.4 & 0.22 & & 19.9 & 3.1 & 0.34 \\
2MASS 1225$-$2739 & 25.5 & 5.2 & 0.16 & & 20.0 & 3.5 & 0.29 \\
2MASS 1237+6526 & 25.3 & 4.1 & 0.24 & & 19.7 & 3.1 & 0.34 \\
Gliese 570D & 25.4 & 5.0 & 0.17 & & 20.0 & 3.8 & 0.27 \\
2MASS 1534$-$2952 & 25.2 & 5.9 & 0.13 & & 19.6 & 4.2 & 0.23 \\
2MASS 1546$-$3325 & 25.4 & 5.0 & 0.17 & & 19.9 & 3.2 & 0.33 \\
2MASS 2356$-$1553 & 25.4 & 4.8 & 0.19 & & 20.0 & 2.9 & 0.36 \\
\enddata
\tablenotetext{a}{S/N = 7 detection limit.} \tablenotetext{b}{Mass
ratio limit derived from Eqn.\ 3.}
\end{deluxetable}

\begin{deluxetable}{cccccccc}
\tabletypesize{\scriptsize}
\tablecaption{Estimated Orbital Parameters.}
\tablewidth{0pt}
\tablehead{
 &  \multicolumn{3}{c}{2MASS 1225$-$2739AB} & &
\multicolumn{3}{c}{2MASS 1534$-$2952AB} \\
 \cline{2-4} \cline{6-8}\noalign{\smallskip}
\colhead{Age (Gyr)} &
\colhead{M (M$_{\sun}$)}  &
\colhead{$a_{sm}$ (AU)\tablenotemark{a}}  &
\colhead{$P$ (yr)}  & &
\colhead{M (M$_{\sun}$)}  &
\colhead{$a_{sm}$ (AU)\tablenotemark{a}}  &
\colhead{$P$ (yr)}   \\
\colhead{(1)} &
\colhead{(2)}  &
\colhead{(3)}  &
\colhead{(4)}  & &
\colhead{(5)}  &
\colhead{(6)}  &
\colhead{(7)}
}
\startdata
 0.5 & 0.017/0.023 & 4.0 & 40 & & 0.027/0.027 & 1.3 & 6.4 \\
 1.0 & 0.024/0.033 & 4.0 & 34 & & 0.035/0.035 & 1.3 & 5.6  \\
 5.0 & 0.04/0.06 & 4.0 & 24 & & 0.065/0.065 & 1.3 & 4.1 \\
\enddata
\tablenotetext{a}{Semimajor axis assuming $<a_{sm}>$ = 1.26$<a>$
\citep{fis92}.}
\end{deluxetable}

\begin{deluxetable}{llccccccc}
\tabletypesize{\scriptsize}
\tablecaption{Field Late-M, L, and T Dwarf Binaries.}
\tablewidth{0pt}
\tablehead{
\colhead{Object} &
\colhead{SpT} &
\colhead{M (M$_{\sun}$)} &
\colhead{$q$} &
\colhead{$a$ ($\arcsec$)} &
\colhead{$a$ (AU)\tablenotemark{a}} &
\colhead{$\Delta$M} &
\colhead{Filter} &
\colhead{Ref.} \\
\colhead{(1)} &
\colhead{(2)} &
\colhead{(3)} &
\colhead{(4)} &
\colhead{(5)} &
\colhead{(6)} &
\colhead{(7)} &
\colhead{(8)} &
\colhead{(9)}
}
\startdata
2MASS 2206$-$2047AB & M8/M8.5 & 0.090/0.088 & 1.0 & 0$\farcs$17 & 4.1 & 0.11 & JHK$^{\prime}$K & 1 \\
Gliese 569BC & M8.5/M9 & 0.069/0.059 & 0.9 & 0$\farcs$08 & 0.90$\pm$0.02 & 0.45 & JK & 2 \\
2MASS 2331$-$0406AB & M8/L3 & 0.091/0.062 & 0.7 & 0$\farcs$57 & 14.4 & 2.6 & JHK$^{\prime}$K & 1 \\
LHS 2397aAB & M8/L7.5 & 0.090/0.068 & 0.8 & 0$\farcs$21 & 3.0 & 4.5 & F814W & 3 \\
2MASS 1426+1557AB & M8.5/L1 & 0.083/0.075 & 0.9 &  0$\farcs$15 & 3.6 & 0.67 & JHK$^{\prime}$K & 1 \\
2MASS 2140+1652AB & M8.5/L0 & 0.087/0.075 & 0.9 & 0$\farcs$16 & 3.7 & 0.75 & JHK$^{\prime}$K & 1 \\
2MASS 0746+2000AB & L0.5/L0.5  & 0.075/0.075\tablenotemark{b} & 1.0 & 0$\farcs$22 & 2.7 & 0.63 & F814W & 4 \\
2MASS 1146+2230AB & L3/L3 & 0.06/0.06\tablenotemark{c}  & 1.0 & 0$\farcs$29 & 7.6 & 0.31 & F814W & 4 \\
 &  &  &  & 0$\farcs$29 & 7.6 & 0.0 & K & 5 \\
Gliese 564BC &  L4/L4  & 0.053/0.053 & 1.0 & 0$\farcs$13 &  2.4 & 0.30 & JHK$_s$ & 6,7 \\
DENIS 1228$-$1159AB & L5/L5  & 0.06/0.06\tablenotemark{c}  & 1.0 & 0$\farcs$28 & 5.1 & 0.22 & F110M & 8,4 \\
 &  & & & 0$\farcs$27 & 4.9 & 0.10 & K & 5 \\
2MASS 0850+1057AB & L6/T?  &  0.05/0.04 & 0.8 & 0$\farcs$16 & 4.4 & 1.3 & F814W & 4 \\
2MASS 0920+3517AB & L6.5/L6.5  & 0.68/0.68 & 1.0 & 0$\farcs$07 & 1.6 & 0.43 & F814W & 4 \\
DENIS 0205$-$1159AB & L7/L7  &  0.05/0.05 & 1.0 & 0$\farcs$51 & 9.2 & 0.0 & K & 5 \\
 &  & & & 0$\farcs$35 & 6.3 & 0.0 & JHKL' & 9 \\
2MASS 1534$-$2952AB & T5.5/T5.5  &  0.035/0.035\tablenotemark{b} & 1.0 & 0$\farcs$07 & 1.0 & 0.5 & F814W & 10 \\
2MASS 1225$-$2739AB & T6/T8  &  0.033/0.024\tablenotemark{b} & 0.7 & 0$\farcs$28 & 3.2 & 1.6 & F814W & 10 \\
\enddata
\tablenotetext{a}{Projected separation, except for Gliese 569BC
whose orbit has been mapped \citep{lan01}.}
\tablenotetext{b}{Assuming and age of 1 Gyr, T$_{eff}$ scale from \citet{me02},
and evolutionary models from \citet{bur97}.}
\tablenotetext{c}{Upper limit on masses based on the detection of 6708 {\AA}
Li absorption \citep{kir99,rei01a}.}
\tablerefs{
(1) \citet{clo02}; (2) \citet{lan01};
(3) \citet{fre02}; (4) \citet{rei01a}; (5) \citet{koe99};
(6) \citet{pot02}; (7) \citet{got02};
(8) \citet{mrt99c}; (9) \citet{leg01};
(10) This paper.}
\end{deluxetable}

\begin{figure}
\epsscale{0.8}
\plotone{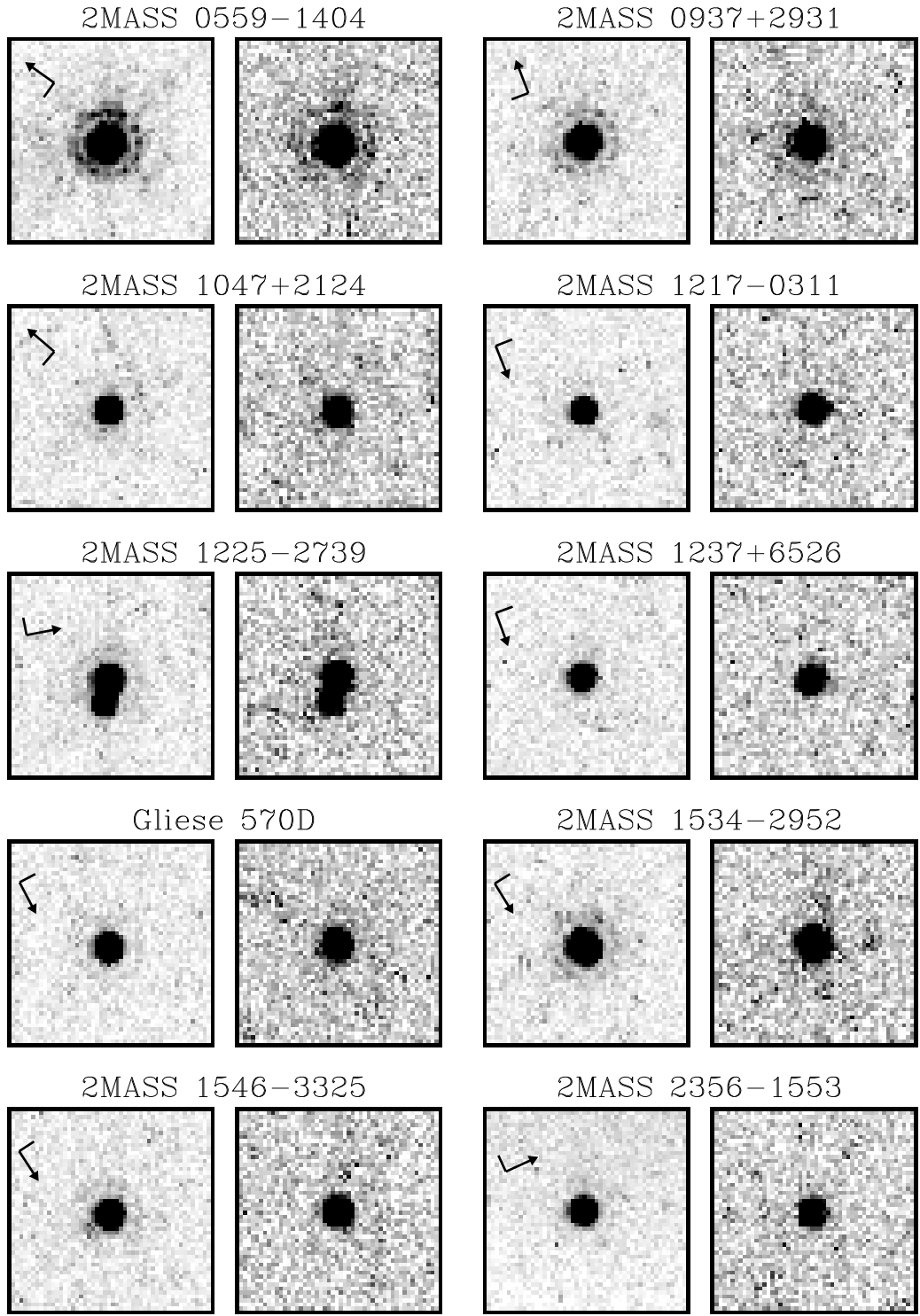} \caption{PC chip images
around each target source.  F814W images are on the left while F1042M
images are on the right.  Each image is $2{\farcs}3$ on a side
with a pixel scale of 0$\farcs$0455.  Image orientations are
indicated by the inset arrows, with the arrowhead pointing North
and orthogonal line pointing East.}
\end{figure}

\begin{figure}
\epsscale{0.8}
\plotone{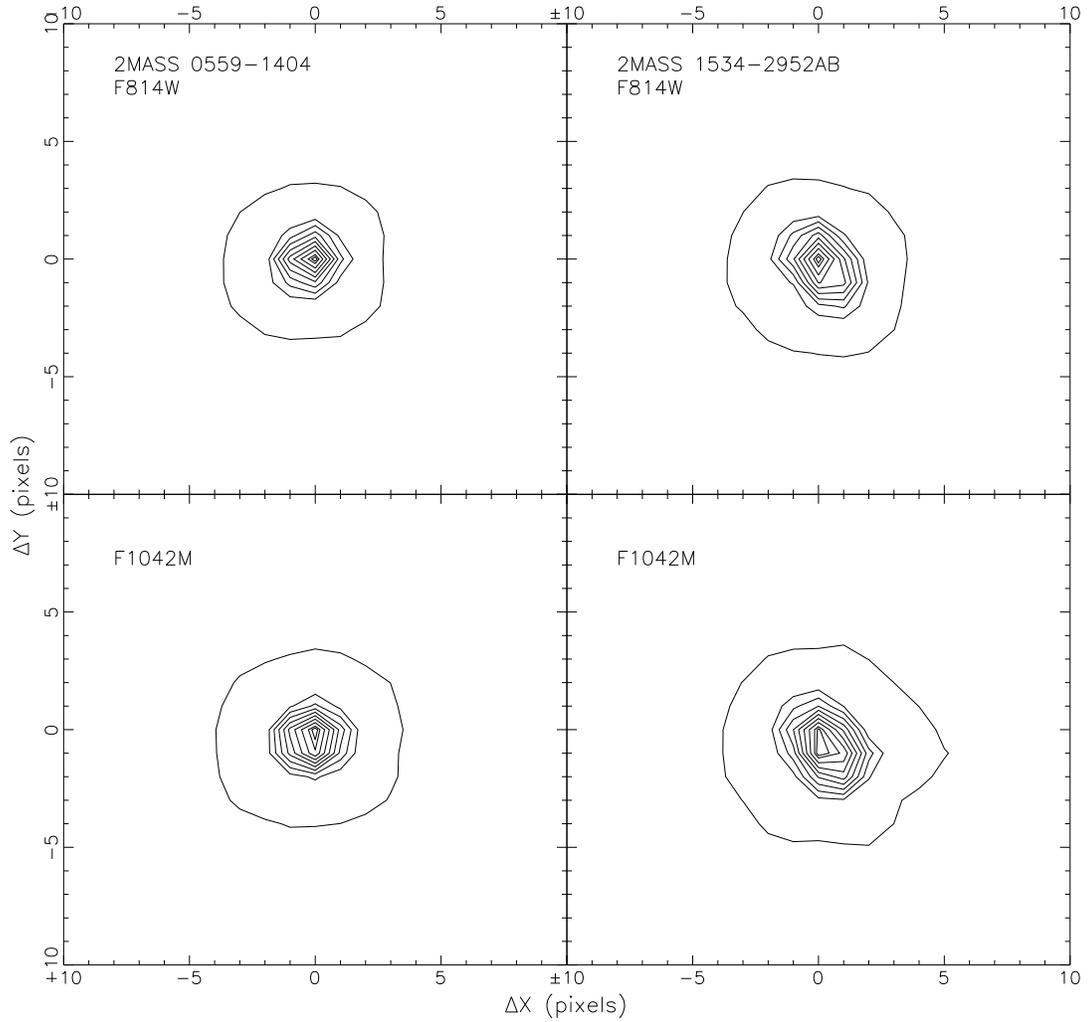} \caption{Contour plots of the F814W (top) and
F1042M (bottom) PC chip
images of 2MASS 0559$-$1404 (left) and 2MASS 1534$-$2952AB (right).
Areas shown are 20 pixels on a side, corresponding to 0$\farcs$92,
and orientations are the same as in Figure 1.  Contour levels of 5\%,
20\%, 30\%, 40\%, 50\%, 60\%, 70\%, 80\%, 90\%, and 95\% of the source
peak are shown.}
\end{figure}

\begin{figure}
\plotone{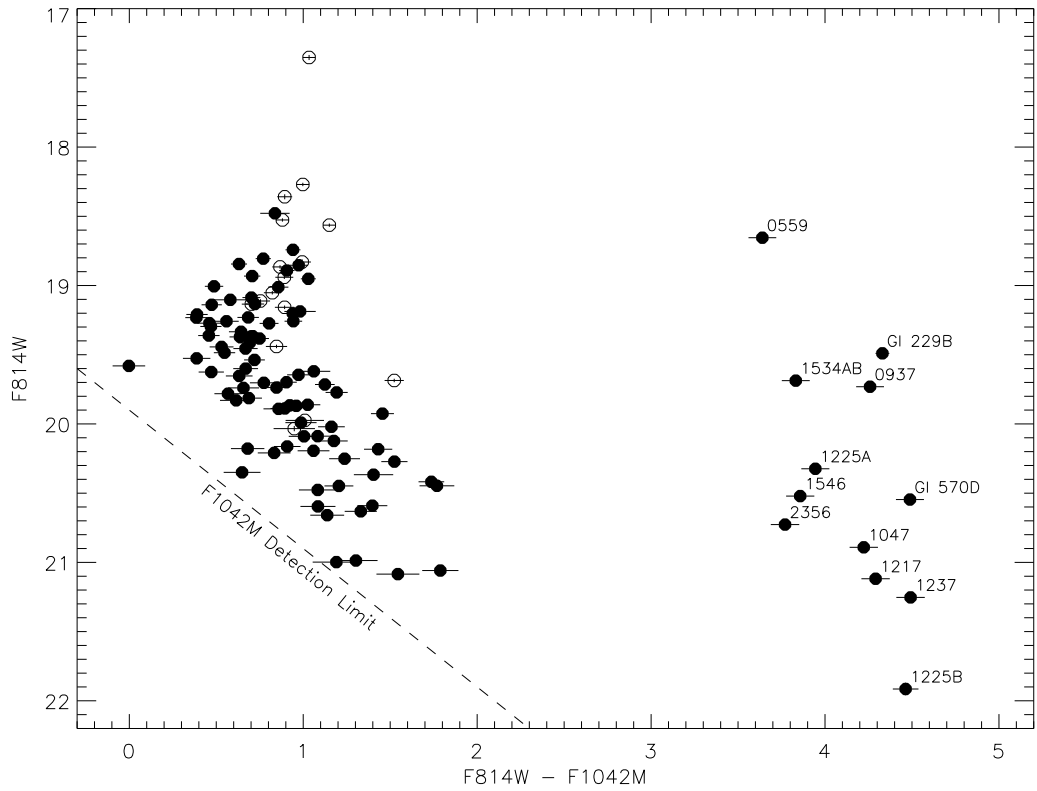} \caption{Color-magnitude diagram of all
sources detected at both F814W and F1042M.  3-pixel aperture
photometry for point sources and target objects (labelled) are
indicated by filled symbols, while 5-pixel aperture photometry for
extended sources (e.g.., galaxies) are indicated by open symbols.
The dashed line indicates the limit for detections at both F814W and
F1042M.}
\end{figure}

\begin{figure}
\plotone{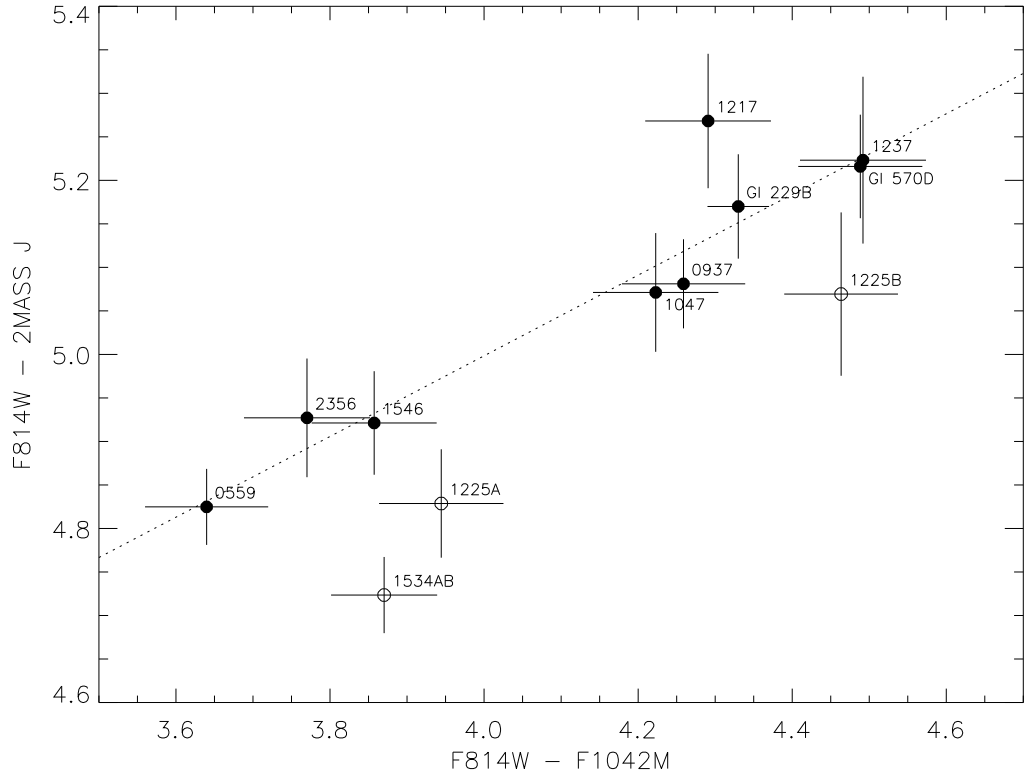} \caption{Optical/near-infrared
color-color diagram for target objects.  2MASS J-band photometry
is used for all target objects.  Individual photometry for the two
components of the 2MASS 1225$-$2739AB system are derived using
Eqn.\ 2 and the combined light magnitude J = 15.22$\pm$0.05.  Data
for Gliese 229B are from \citet{gol98} and \citet{leg99}.  A
straight-line fit to all single sources (excluding Gliese 229B) is
indicated by the dashed line. }
\end{figure}

\begin{figure}
\plottwo{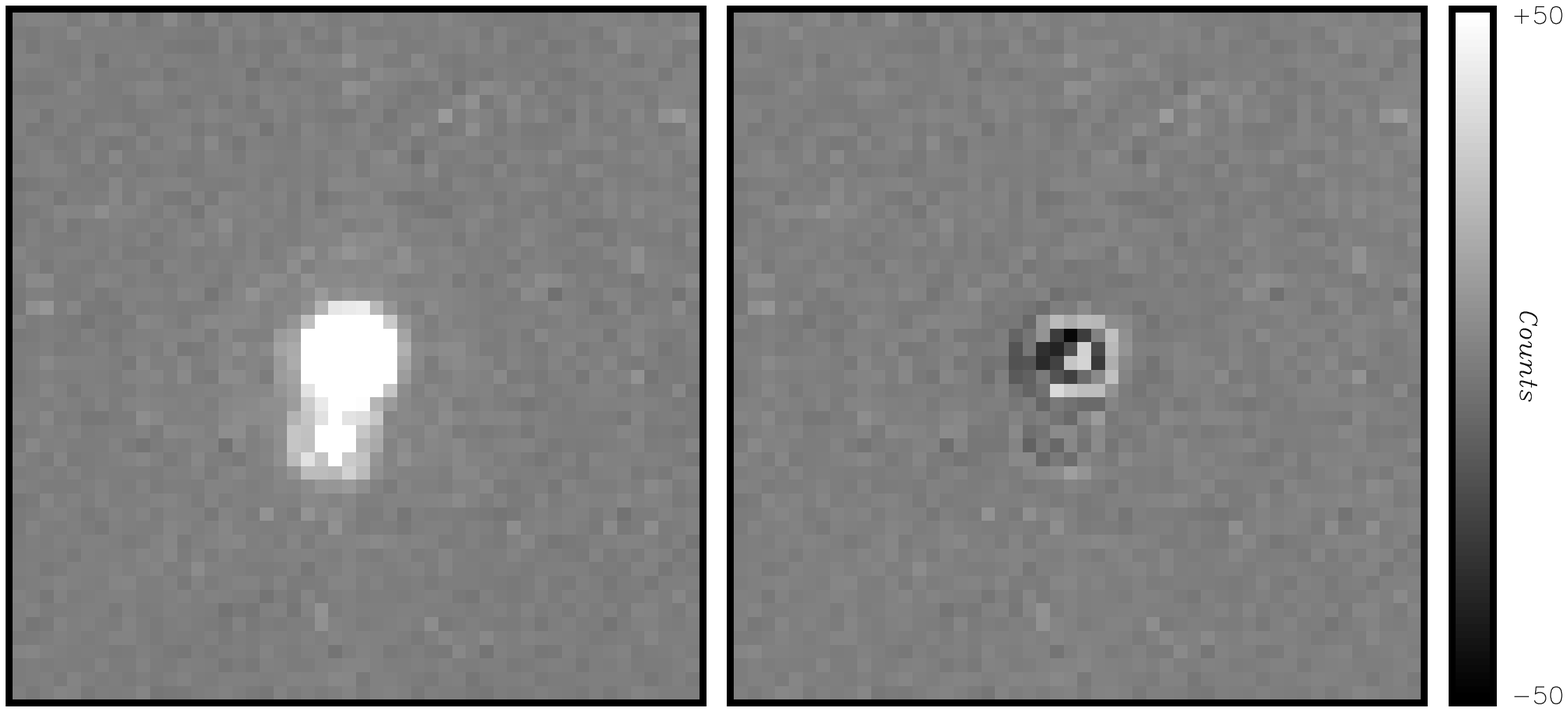}{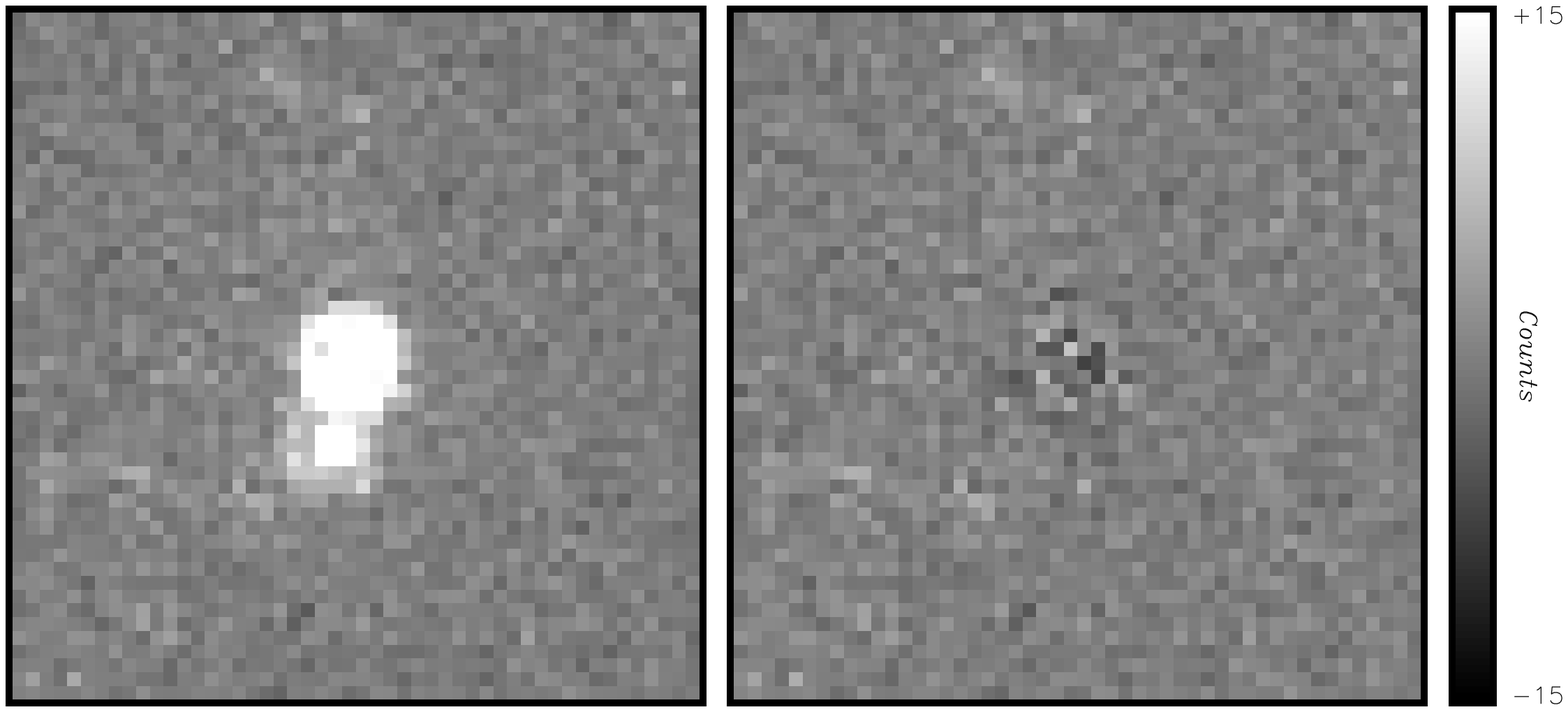} \plotone{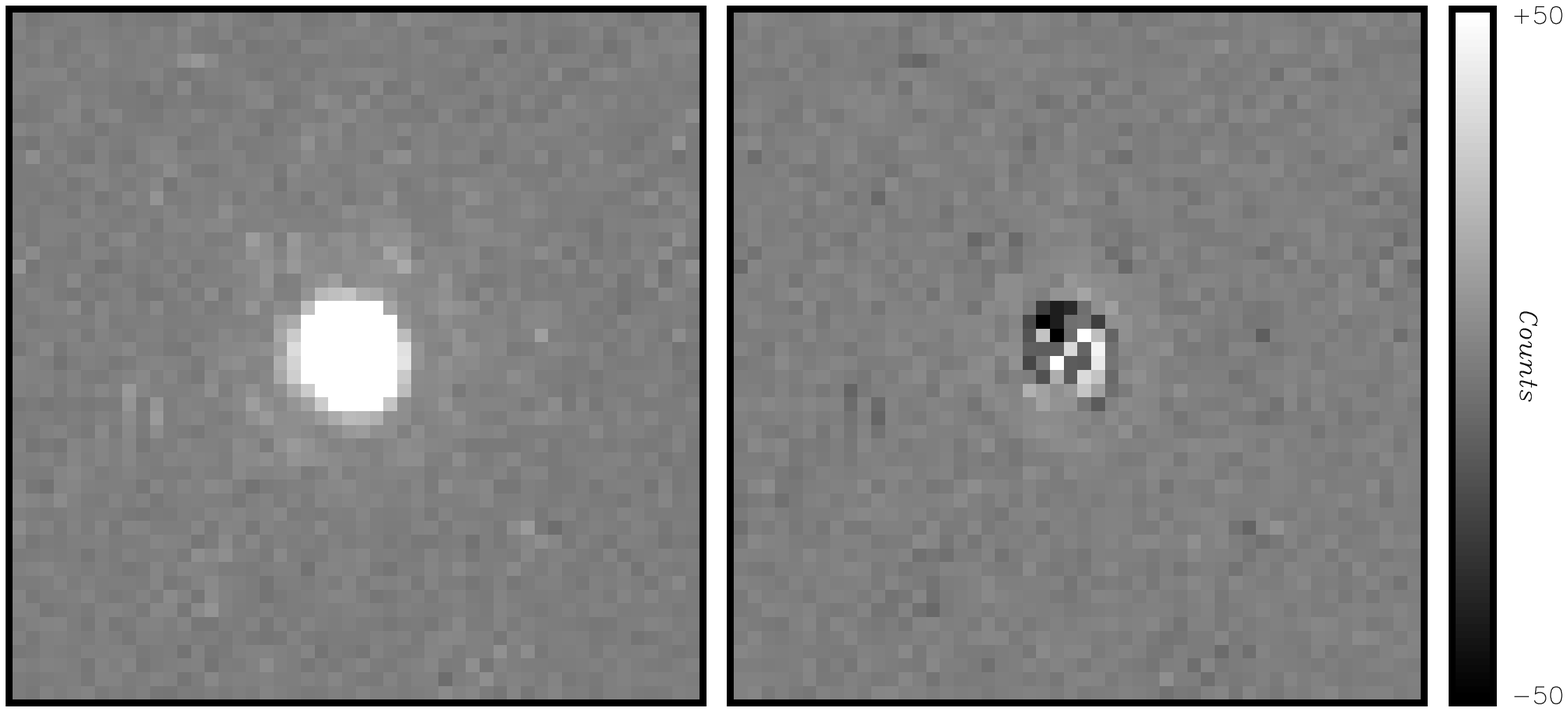}
\plotone{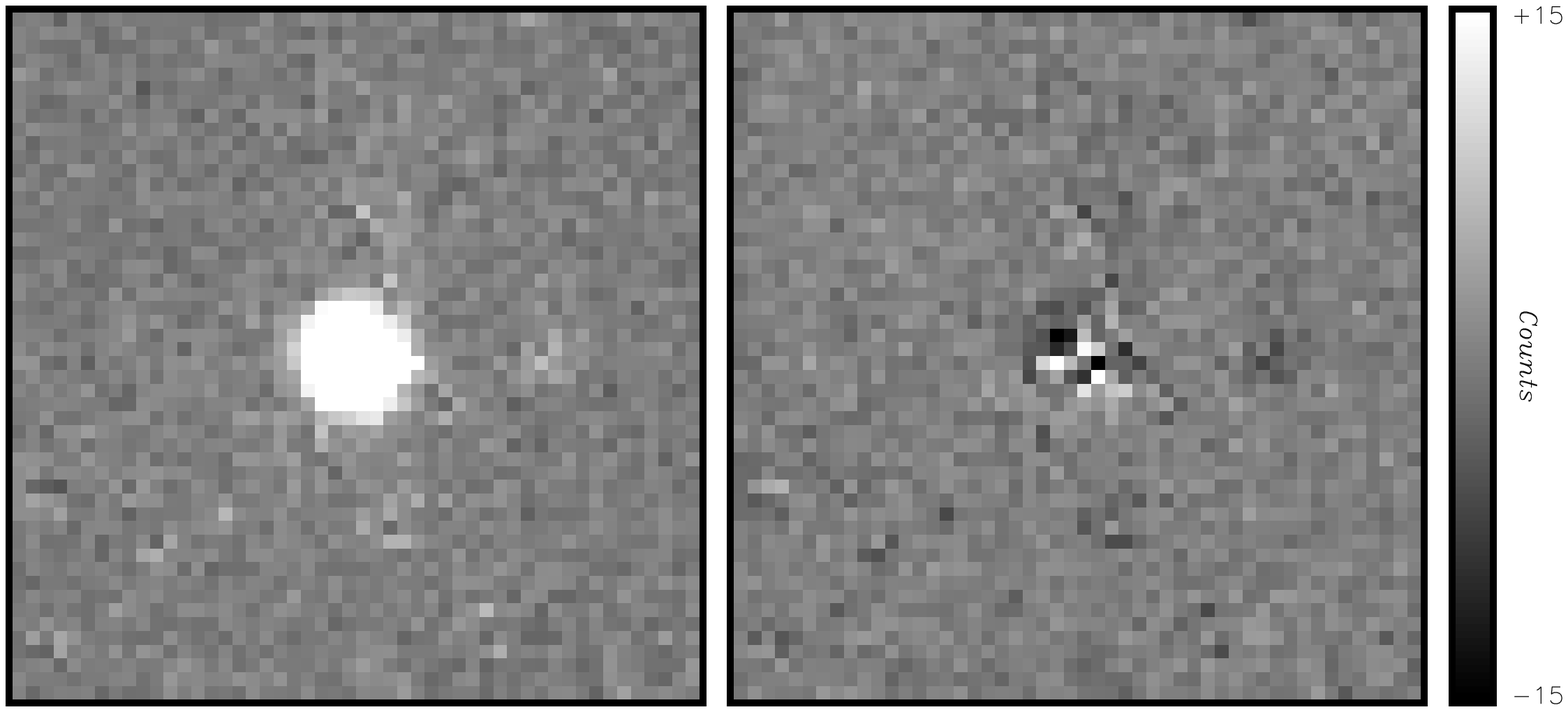} \caption{PSF subtraction for 2MASS
1225$-$2739AB (top) and 2MASS 1534$-$2952AB (bottom).  Both F814W
(left) and F1042M (right) images are shown.  The first image for
each set shows the original PC image, while the second shows the
residual image after subtracting the PSF model.  Color scales are
given for each set. }
\end{figure}

\begin{figure}
\plottwo{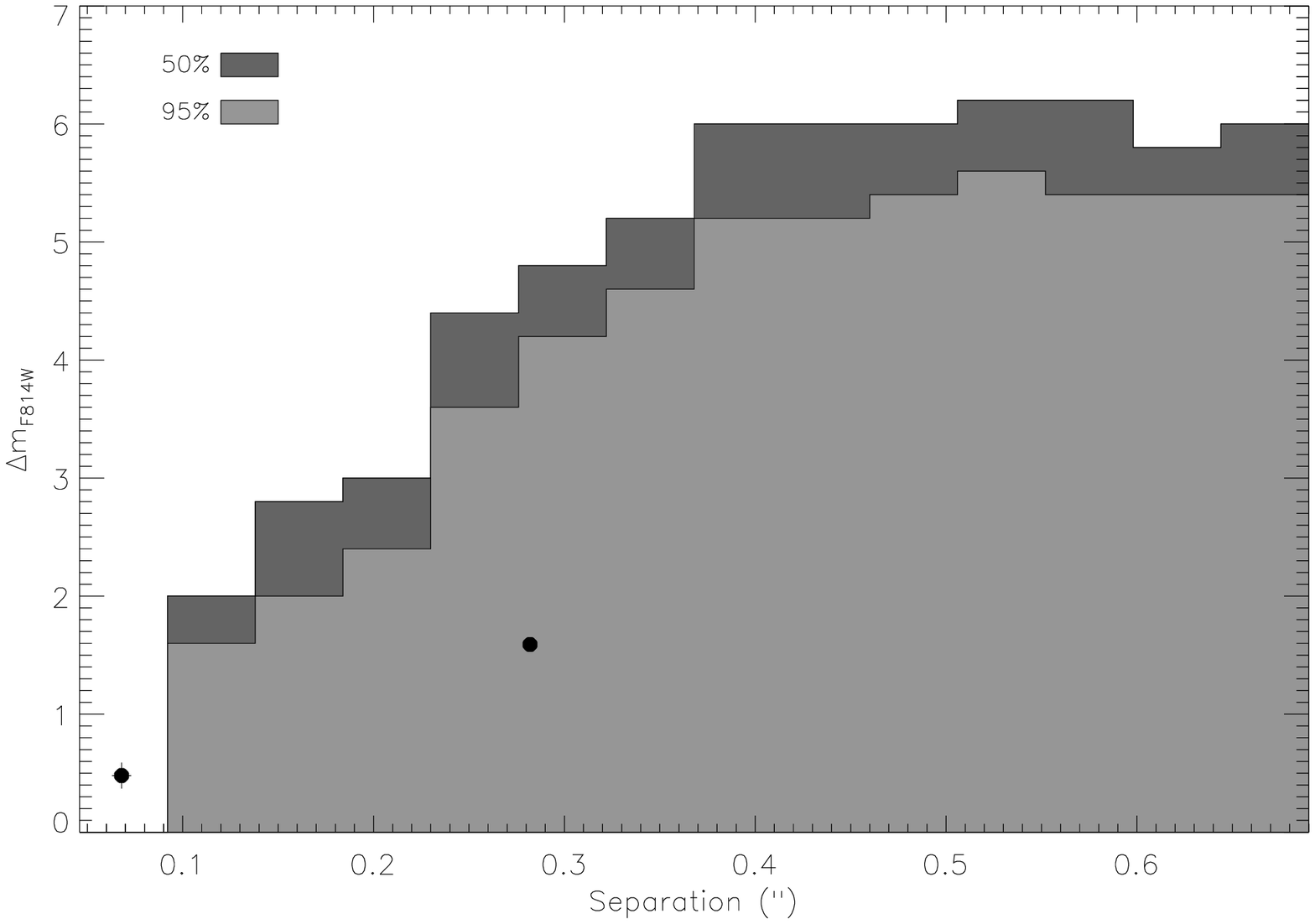}{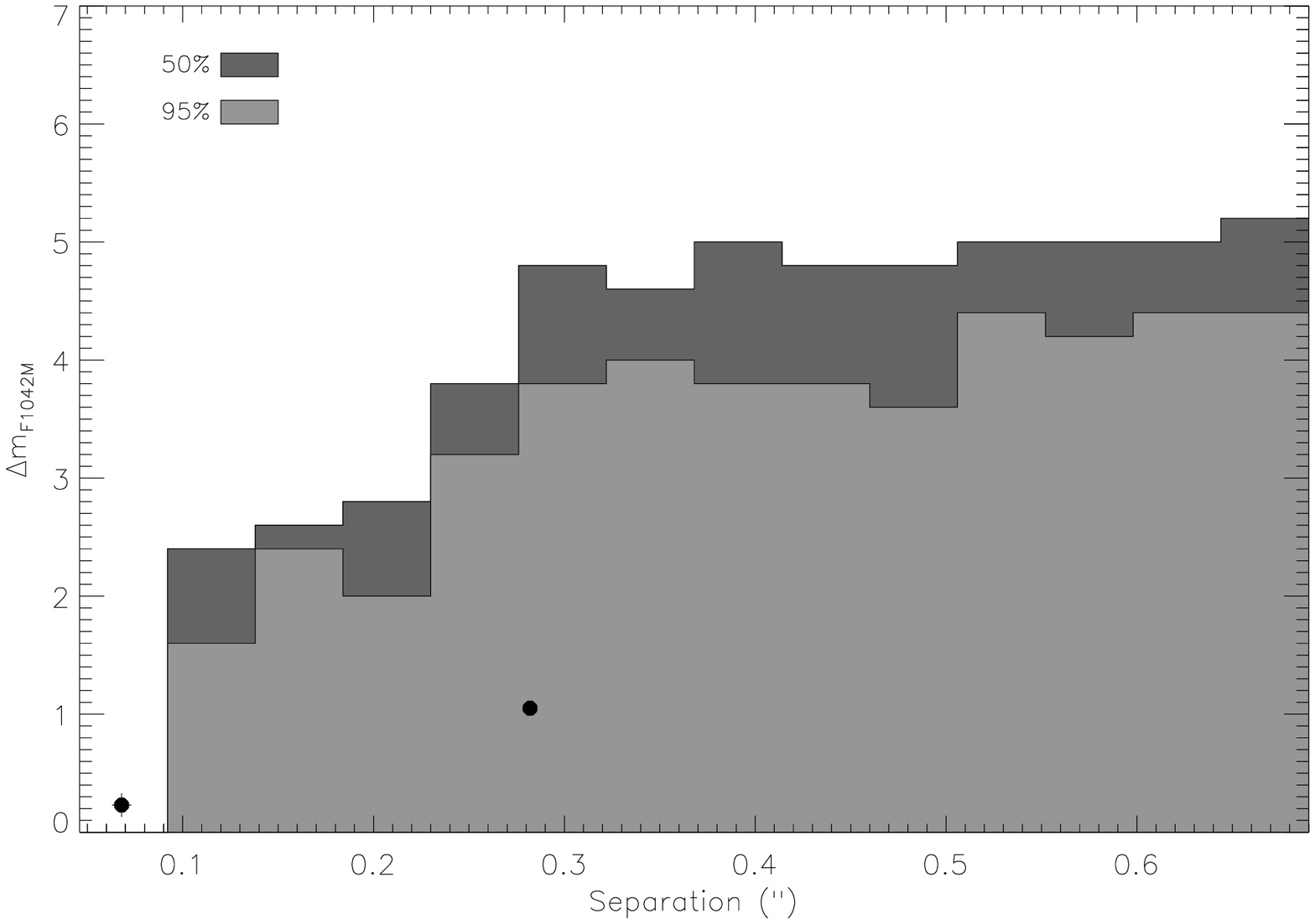}
\caption{Completeness limits for companions around 2MASS
0559$-$1404 in the F814W (left) and F1042M (right) filters.  The
light grey histogram gives the 95\% recovery limit; i.e., the
limiting flux ratio at which 95\% of the simulated binaries were
accurately extracted by our fitting algorithm at each separation.
The 50\% recovery limit is shown in dark grey.  The separations
and flux ratios of 2MASS 1225$-$2739AB and 2MASS 1534$-$2952AB are
indicated by solid circles.}
\end{figure}

\begin{figure}
\plotone{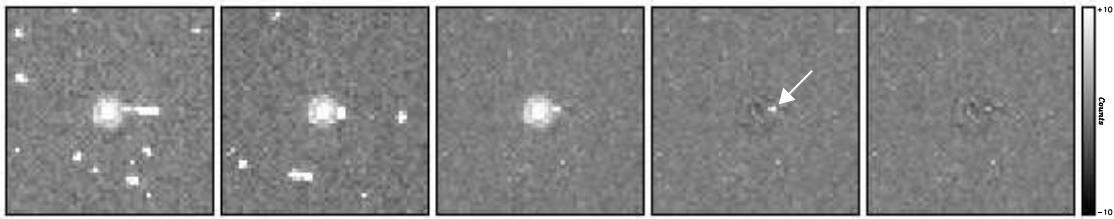} \caption{A possible companion to
2MASS 1217$-$0311.  The first two panels show the raw F1042M image
frames centered on 2MASS 1217$-$0311 prior to cosmic ray
correction; the third panel shows the corrected and combined image
frame; the fourth panel shows the single source PSF-subtracted image,
with the putative companion indicated by an arrow; the fifth panel
shows the residual image after subtraction of two PSFs.
All images are 2$\farcs$3 on a side, and color scale
is indicated on the right. }
\end{figure}

\begin{figure}
\plotone{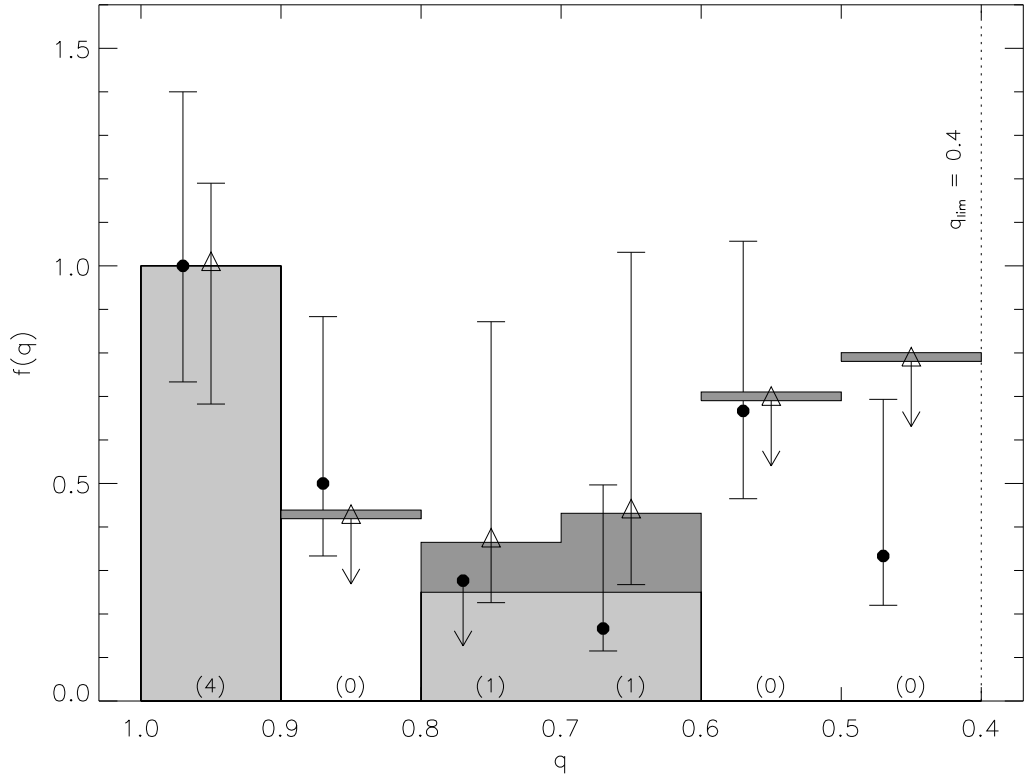} \caption{Combined mass ratio distribution for
T dwarf binaries in this sample and L dwarf binaries in
\citet{rei01a}. Individual mass ratios are listed in Table 7.  The
light grey histogram shows the observed distribution normalized to
$f(q=1)$ = 1; number counts are shown in parentheses at the bottom
of each bin.  The observed distribution overlaps a bias-corrected
distribution indicated by the dark grey histogram and triangles
with 1$\sigma$ uncertainties and upper limits (arrows).
The slightly offset filled circles denote the mass
ratio distribution (with the same normalization) of M dwarf
binaries in the 8 pc sample \citep{rei97}, with uncertainties
computed as described in the Appendix (an upper limit for $0.7 < q
< 0.8$ is indicated by the downward arrow). The mass ratio limit,
$q_{lim} \approx 0.4$, for the combined L and T sample for $a
\gtrsim 4-5$ AU is indicated by the short-dashed line.}
\end{figure}

\begin{figure}
\plotone{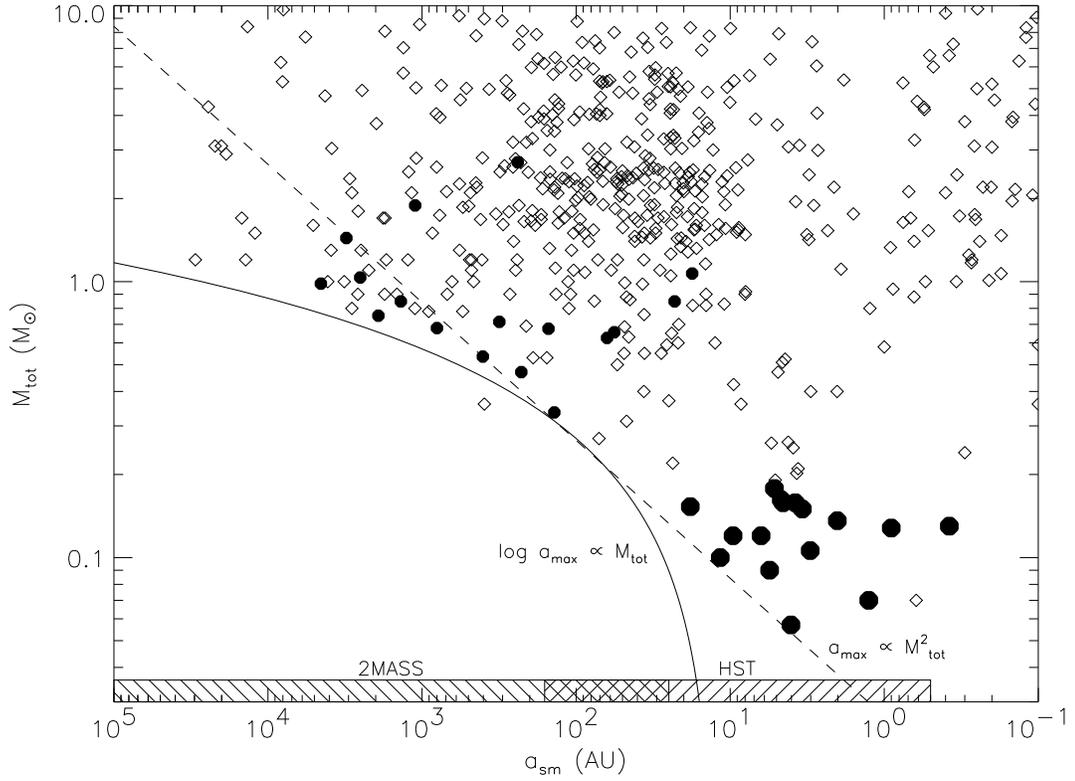} \caption{Total mass (M$_{tot}$) versus
separation ($a$) for star-star (open diamonds), star-brown dwarf
(small filled circles) and late-type dwarf (primary star later
than M8; large filled circles) binaries.  The maximum separation
for the more massive systems (M$_{tot}$ $\gtrsim 0.4$ M$_{\sun}$)
appears to be limited by $\log{a_{max}} \propto$ M$_{tot}$
\citep[solid line]{rei01a}, while the low-mass binary envelope
(dashed line) appears to follow $a_{max} \propto$ M$_{tot}^2$.
Resolvable separations for L and T dwarfs (typical distance of 20
and 10 pc, respectively) for 2MASS and HST are indicated along the
bottom of the figure.}
\end{figure}

\begin{figure}
\plotone{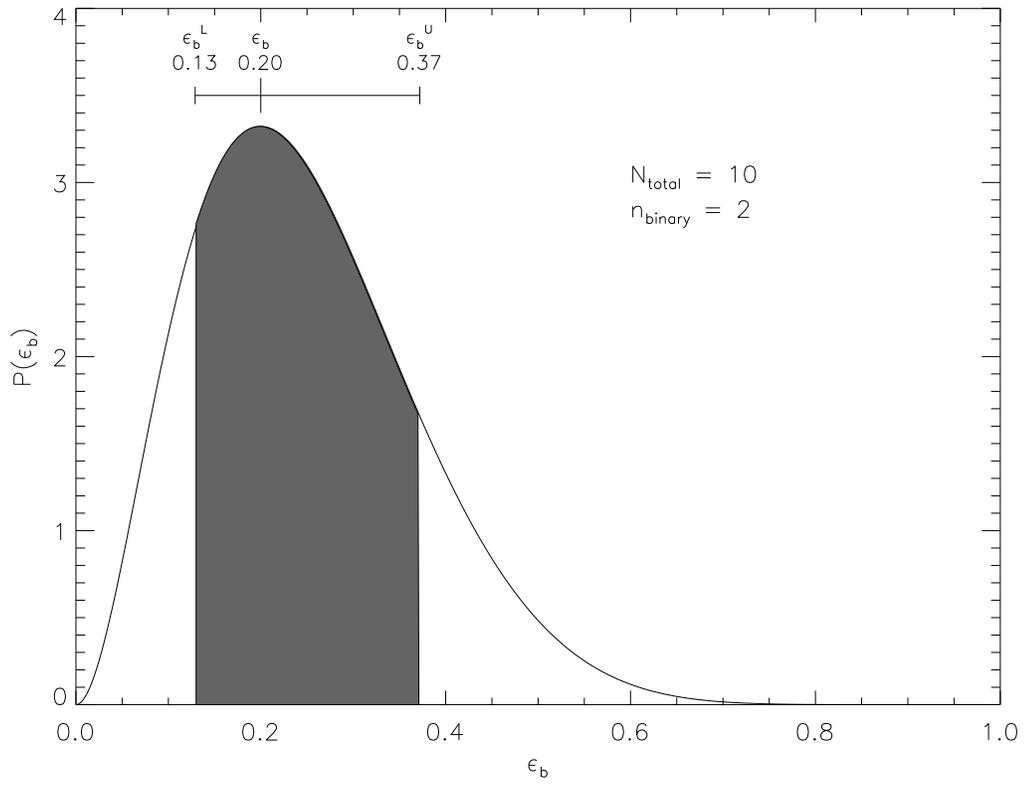} \caption{Probability distribution for
${\epsilon}_b$ constructed for a sample size $N = 10$ and number
of binaries $n = 2$.  The shaded region gives the ${\pm}1{\sigma}$
range of acceptable values, whose limits are defined in the
Appendix.  The integrated probability in this region, 68\%, is
equivalent to 1$\sigma$ Gaussian limits.}
\end{figure}

\end{document}